\documentstyle{paper}

\def\txstrut{ \vrule height 15pt depth 10pt width 0pt }
\def\frac#1#2{{\displaystyle{#1}\over\displaystyle{#2}}}

\def\stackunder#1#2{\mathrel{\mathop{#2}\limits_{#1}}}

\catcode`\@=11

\let\DOTSI\relax
\def\RIfM@{\relax\ifmmode}
\def\FN@{\futurelet\next}
\newcount\intno@
\def\iint{\DOTSI\intno@\tw@\FN@\ints@}
\def\iiint{\DOTSI\intno@\thr@@\FN@\ints@}
\def\iiiint{\DOTSI\intno@4 \FN@\ints@}
\def\idotsint{\DOTSI\intno@\z@\FN@\ints@}
\def\ints@{\findlimits@\ints@@}
\newif\iflimtoken@
\newif\iflimits@
\def\findlimits@{\limtoken@true\ifx\next\limits\limits@true
 \else\ifx\next\nolimits\limits@false\else
 \limtoken@false\ifx\ilimits@\nolimits\limits@false\else
 \ifinner\limits@false\else\limits@true\fi\fi\fi\fi}
\def\multint@{\int\ifnum\intno@=\z@\intdots@                                
 \else\intkern@\fi                                                          
 \ifnum\intno@>\tw@\int\intkern@\fi                                         
 \ifnum\intno@>\thr@@\int\intkern@\fi                                       
 \int}                                                                      
\def\multintlimits@{\intop\ifnum\intno@=\z@\intdots@\else\intkern@\fi
 \ifnum\intno@>\tw@\intop\intkern@\fi
 \ifnum\intno@>\thr@@\intop\intkern@\fi\intop}
\def\intic@{\mathchoice{\hskip.5em}{\hskip.4em}{\hskip.4em}{\hskip.4em}}
\def\negintic@{\mathchoice
 {\hskip-.5em}{\hskip-.4em}{\hskip-.4em}{\hskip-.4em}}
\def\ints@@{\iflimtoken@                                                    
 \def\ints@@@{\iflimits@\negintic@\mathop{\intic@\multintlimits@}\limits    
  \else\multint@\nolimits\fi                                                
  \eat@}                                                                    
 \else                                                                      
 \def\ints@@@{\iflimits@\negintic@
  \mathop{\intic@\multintlimits@}\limits\else
  \multint@\nolimits\fi}\fi\ints@@@}
\def\intkern@{\mathchoice{\!\!\!}{\!\!}{\!\!}{\!\!}}
\def\plaincdots@{\mathinner{\cdotp\cdotp\cdotp}}
\def\intdots@{\mathchoice{\plaincdots@}
 {{\cdotp}\mkern1.5mu{\cdotp}\mkern1.5mu{\cdotp}}
 {{\cdotp}\mkern1mu{\cdotp}\mkern1mu{\cdotp}}
 {{\cdotp}\mkern1mu{\cdotp}\mkern1mu{\cdotp}}}

\newif\iffirstchoice@
\firstchoice@true
\def\textfonti{\the\textfont\@ne}
\def\textfontii{\the\textfont\tw@}
\def\text{\RIfM@\expandafter\text@\else\expandafter\text@@\fi}
\def\text@@#1{\leavevmode\hbox{#1}}
\def\text@#1{\mathchoice
 {\hbox{\everymath{\displaystyle}\def\textfonti{\the\textfont\@ne}%
  \def\textfontii{\the\textfont\tw@}\textdef@@ T#1}}
 {\hbox{\firstchoice@false
  \everymath{\textstyle}\def\textfonti{\the\textfont\@ne}%
  \def\textfontii{\the\textfont\tw@}\textdef@@ T#1}}
 {\hbox{\firstchoice@false
  \everymath{\scriptstyle}\def\textfonti{\the\scriptfont\@ne}%
  \def\textfontii{\the\scriptfont\tw@}\textdef@@ S\rm#1}}
 {\hbox{\firstchoice@false
  \everymath{\scriptscriptstyle}\def\textfonti
  {\the\scriptscriptfont\@ne}%
  \def\textfontii{\the\scriptscriptfont\tw@}\textdef@@ s\rm#1}}}
\def\textdef@@#1{\textdef@#1\rm\textdef@#1\bf\textdef@#1\sl\textdef@#1\it}
\def\DN@{\def\next@}
\def\eat@#1{}
\def\textdef@#1#2{%
 \DN@{\csname\expandafter\eat@\string#2fam\endcsname}%
 \if S#1\edef#2{\the\scriptfont\next@\relax}%
 \else\if s#1\edef#2{\the\scriptscriptfont\next@\relax}%
 \else\edef#2{\the\textfont\next@\relax}\fi\fi}

\def\Let@{\relax\iffalse{\fi\let\\=\cr\iffalse}\fi}
\def\vspace@{\def\vspace##1{\crcr\noalign{\vskip##1\relax}}}
\def\multilimits@{\bgroup\vspace@\Let@
 \baselineskip\fontdimen10 \scriptfont\tw@
 \advance\baselineskip\fontdimen12 \scriptfont\tw@
 \lineskip\thr@@\fontdimen8 \scriptfont\thr@@
 \lineskiplimit\lineskip
 \vbox\bgroup\ialign\bgroup\hfil$\m@th\scriptstyle{##}$\hfil\crcr}
\def\Sb{_\multilimits@}
\def\endSb{\crcr\egroup\egroup\egroup}
\def\Sp{^\multilimits@}

\newdimen\ex@
\def\rightarrowfill@#1{$#1\m@th\mathord-\mkern-6mu\cleaders
 \hbox{$#1\mkern-2mu\mathord-\mkern-2mu$}\hfill
 \mkern-6mu\mathord\rightarrow$}
\def\leftarrowfill@#1{$#1\m@th\mathord\leftarrow\mkern-6mu\cleaders
 \hbox{$#1\mkern-2mu\mathord-\mkern-2mu$}\hfill\mkern-6mu\mathord-$}
\def\leftrightarrowfill@#1{$#1\m@th\mathord\leftarrow\mkern-6mu\cleaders
 \hbox{$#1\mkern-2mu\mathord-\mkern-2mu$}\hfill
 \mkern-6mu\mathord\rightarrow$}
\def\overrightarrow{\mathpalette\overrightarrow@}
\def\overrightarrow@#1#2{\vbox{\ialign{##\crcr\rightarrowfill@#1\crcr
 \noalign{\kern-\ex@\nointerlineskip}$\m@th\hfil#1#2\hfil$\crcr}}}

\def\overleftarrow{\mathpalette\overleftarrow@}
\def\overleftarrow@#1#2{\vbox{\ialign{##\crcr\leftarrowfill@#1\crcr
 \noalign{\kern-\ex@\nointerlineskip}$\m@th\hfil#1#2\hfil$\crcr}}}
\def\overleftrightarrow{\mathpalette\overleftrightarrow@}
\def\overleftrightarrow@#1#2{\vbox{\ialign{##\crcr\leftrightarrowfill@#1\crcr
 \noalign{\kern-\ex@\nointerlineskip}$\m@th\hfil#1#2\hfil$\crcr}}}
\def\underrightarrow{\mathpalette\underrightarrow@}
\def\underrightarrow@#1#2{\vtop{\ialign{##\crcr$\m@th\hfil#1#2\hfil$\crcr
 \noalign{\nointerlineskip}\rightarrowfill@#1\crcr}}}

\def\underleftarrow{\mathpalette\underleftarrow@}
\def\underleftarrow@#1#2{\vtop{\ialign{##\crcr$\m@th\hfil#1#2\hfil$\crcr
 \noalign{\nointerlineskip}\leftarrowfill@#1\crcr}}}
\def\underleftrightarrow{\mathpalette\underleftrightarrow@}
\def\underleftrightarrow@#1#2{\vtop{\ialign{##\crcr$\m@th\hfil#1#2\hfil$\crcr
 \noalign{\nointerlineskip}\leftrightarrowfill@#1\crcr}}}

\catcode`\@=\active

\def\frac#1#2{{#1 \over #2}}

\def\stackunder#1#2{\mathrel{\mathop{#2}\limits_{#1}}}

\input{epsf.tex}
\def\FiguresIn#1{\def\@Figures{#1}}

\newcount\GRAPHICSTYPE
\def\graffile#1#2#3#4{\leavevmode\raise -#4 \hbox{%
\raise #3 \hbox{\rule{0.003in}{0.003in}\special{#1}}}%
{\raise -#4 \hbox to #2 {\vrule height#3 width0in depth0in\hfil}}%
}

\def\draftbox#1#2#3#4{\leavevmode\raise -#4 \hbox{\frame{\rlap{\protect\tiny #1}%
\hbox to #2{\vrule height#3 width0in depth0in\hfil}}}}

\newcount\draft
\def\GRAPHIC#1#2#3#4#5{\ifnum\draft=1 \draftbox{#2}{#3}{#4}{#5}\else%
\graffile{#1}{#3}{#4}{#5}\fi}

\def\addtoLaTeXparams#1{\edef\LaTeXparams{\LaTeXparams #1}}

\def\doFRAMEparams#1{\readFRAMEparams#1\end}
\def\readFRAMEparams#1{%
\ifx#1\end%
\let\next=\relax%
\else%
\ifx#1i%
\dispkind=0%
\fi%
\ifx#1d%
\dispkind=1%
\fi%
\ifx#1f%
\dispkind=2%
\fi%
\ifx#1t%
\addtoLaTeXparams{t}%
\fi%
\ifx#1b%
\addtoLaTeXparams{b}%
\fi%
\ifx#1p%
\addtoLaTeXparams{p}%
\fi%
\ifx#1h%
\addtoLaTeXparams{h}%
\fi%
\let\next=\readFRAMEparams%
\fi%
\next%
}

\def\IFRAME#1#2#3#4#5{\GRAPHIC{#5}{#4}{#1}{#2}{#3}}

\def\DFRAME#1#2#3#4{
  \begin{center}
    \GRAPHIC{#4}{#3}{#1}{#2}{0in} 
  \end{center}
}

\def\FFRAME#1#2#3#4#5#6#7{
  \begin{figure}[#1]
    \begin{center}
      \leavevmode\epsffile{\@Figures#7}
    \end{center}
    \caption{\label{#5}#4}
  \end{figure}
}

\def\FRAME#1#2#3#4#5#6#7#8{%
\newcount\dispkind%
\def\LaTeXparams{}%
\dispkind=0%
\def\LaTeXparams{}%
\doFRAMEparams{#1}%
\ifnum\dispkind=0%
\IFRAME{#2}{#3}{#4}{#7}{#8}%
\else
  \ifnum\dispkind=1
    \DFRAME{#2}{#3}{#7}{#8}
  \else
    \ifnum\dispkind=2
      \FFRAME{\LaTeXparams}{#2}{#3}{#5}{#6}{#7}{#8}
    \fi
  \fi
\fi
}

\catcode`\@=11

\long\def\QQQ#1#2{}
\def\QTP#1{}
\long\def\QQA#1#2{}

\def\EXPAND#1[#2]#3{}
\def\NOEXPAND#1[#2]#3{}

\def\LaTeXparent#1{}

\@ifundefined{INDEX}{\def\INDEX#1#2{}{}}{}
\@ifundefined{SUBINDEX}{\def\SUBINDEX#1#2#3{}{}{}}{}
\def\initial#1{\bigbreak{\raggedright\large\bf #1}\kern 2pt\penalty3000}

\@ifundefined{abstract}{%
\def\abstract{\if@twocolumn
\section*{Abstract (Not appropriate in this style!)}
\else \small 
\begin{center}
{\bf Abstract\vspace{-.5em}\vspace{0pt}} 
\end{center}
\quotation 
\fi}}{}
\@ifundefined{endabstract}{%
\def\endabstract{\if@twocolumn\else\endquotation\fi}}{}
\@ifundefined{maketitle}{\def\maketitle#1{}}{}
\@ifundefined{affiliation}{\def\affiliation#1{}}{}
\@ifundefined{proof}{}{}
\@ifundefined{newfield}{\def\newfield#1#2{}}{}
\@ifundefined{chapter}{\def\chapter#1{\par(Chapter head:)#1\par }}{}
\@ifundefined{part}{\def\part#1{\par(Part head:)#1\par }}{}
\@ifundefined{section}{\def\part#1{\par(Section head:)#1\par }}{}
\@ifundefined{subsection}{\def\part#1{\par(Subsection head:)#1\par }}{}
\@ifundefined{subsubsection}{\def\part#1{\par(Subsubsection head:)#1\par }}{}
\@ifundefined{paragraph}{\def\part#1{\par(Subsubsubsection head:)#1\par }}{}
\@ifundefined{subparagraph}{\def\part#1{\par(Subsubsubsubsection head:)#1\par }}{}

\newdimen\theight
\def \Column{%
             \vadjust{\setbox0=\hbox{\scriptsize\quad\quad tcol}%
             \theight=\ht0
             \advance\theight by \dp0    \advance\theight by \lineskip
             \kern -\theight \vbox to \theight{\rightline{\rlap{\box0}}%
             \vss}%
             }}%

\def\qed{\ifhmode\unskip\nobreak\fi\ifmmode\ifinner\else\hskip5\p@\fi\fi
 \hbox{\hskip5\p@\vrule width4\p@ height6\p@ depth1.5\p@\hskip\p@}}
\catcode`@=12 

\makeatletter

\def\FiguresIn#1{\def\@Figures{#1}}
\def\GetFigure#1{\begin{center}\leavevmode\epsffile{\@Figures#1}\end{center}}
\def\PutFigure#1{\begin{array}{c}{\epsffile{\@Figures#1}}\end{array}}
\def\putFigure#1{\begin{array}{c}\framebox{\epsffile{\@Figures#1}}\end{array}}

\def\alphas{\alpha_{s}}

\def\ie{i \varepsilon}
\newskip\humongous \humongous=0pt plus 1000pt minus 1000pt

\newif\ifdtup

\def\picture #1 by #2 (#3){\vbox to #2{\hrule width #1 height
  0pt depth 0pt\vfill\special{picture #3}}}
\def\scaledpicture #1 by #2 (#3 scaled #4){{\dimen0=#1 \dimen1=#2
  \divide\dimen0 by 1000 \multiply\dimen0 by #4
  \divide\dimen1 by 1000 \multiply\dimen1 by #4
  \picture \dimen0 by \dimen1 (#3 scaled #4)}}
\def\tstrut{ \vrule height 15pt depth 5pt width 0pt }
\def\ttstrut{\vrule height 20pt depth 20pt width 0pt }

\def\ch{{\rm cosh}}
\def\sh{{\rm sinh}}

\newcommand{\App}[1] {Appendix~\ref{app:#1}}

\newcommand{\Eq}[1]  {Eq.~(\ref{eq:#1})}
\newcommand{\Eqs}[2] {Eqs.~(\ref{eq:#1})--(\ref{eq:#2})}
\newcommand{\Fig}[1] {Fig.~\ref{fig:#1}}

\newcommand{\rf}[1]  {${}^{\,\cite{#1}}$}

\newcommand{\Sec}[1] {Section~\ref{sec:#1}}
\newcommand{\Tbl}[1] {Table~\ref{tab:#1}}

\newcommand{\cf}  {{\it cf.\ }}

\newcommand{\et}{\eta}

\newcommand{\la}{\lambda}

\def\ie{{\it i.e.}}

\def\cf{{\it cf.}}

\begin{document}


\def\macgraphsize#1#2#3{\vbox to #2{\hsize #1\relax%
\hrule height 0pt depth 0pt width 0pt\vfill%
\special{postscriptfile #3}}}


\newdimen\xhsize

\def\doit#1#2{   %
\xhsize=\hsize   %
\multiply\xhsize by #1 \divide\xhsize by #2   %
\epsfxsize=\xhsize }

\def\xfiglhProc{
\begin{figure}[tbh]
 \epsfxsize=\hsize
  \epsfbox{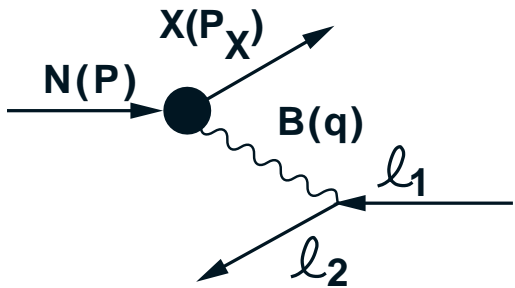}
\caption{ 
The general lepton-hadron scattering process:  $N(P) + \ell_1 \rightarrow
X(P_X) + \ell_2$ via the exchange of  a vector boson, $B(q)$. The lepton
momenta are  $\ell_{i}$ while the initial and final hadronic  momenta are
$P$ and $P_{X}$, respectively. } 
  \label{fig:lhProc}
\end{figure}
}
\def\xfigFactThm{
\begin{figure}[bht]
 \epsfxsize=\hsize
  \epsfbox{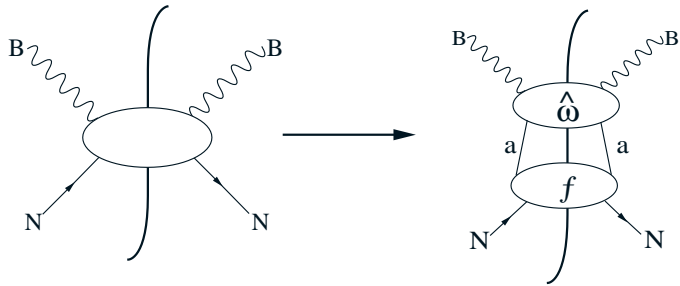}
\caption{
Pictorial representation of the factorization theorem for the hadron
structure functions for inclusive deeply inelastic scattering.
The process on the left is  $N(P) + B(q) \rightarrow X(P_X)$, and the
factorized process on the right is  $N(P) \rightarrow a(k_1)$
(represented by the parton distribution function, $f_N^a$) with the
successive hard scattering interaction   $a(k_1) + B(q)$
(represented by  $\omega_{\mu\nu}^a$). The vertical  lines
indicate an inclusive sum over the final states, $X(P_X)$.
} 
  \label{fig:FactThm}
\end{figure}
}
\def\xfigBorn{
\begin{figure}[bh]
 \epsfxsize=\hsize
  \epsfbox{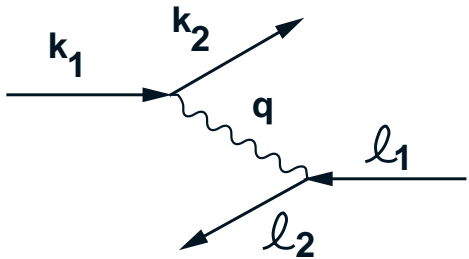}
\caption{ Leading order hard-scattering amplitude 
for heavy quark production.} 
  \label{fig:Born}
\end{figure}
}
\def\xfigSRdy{
\begin{figure}[tbh]
 \epsfxsize=\hsize
  \epsfbox{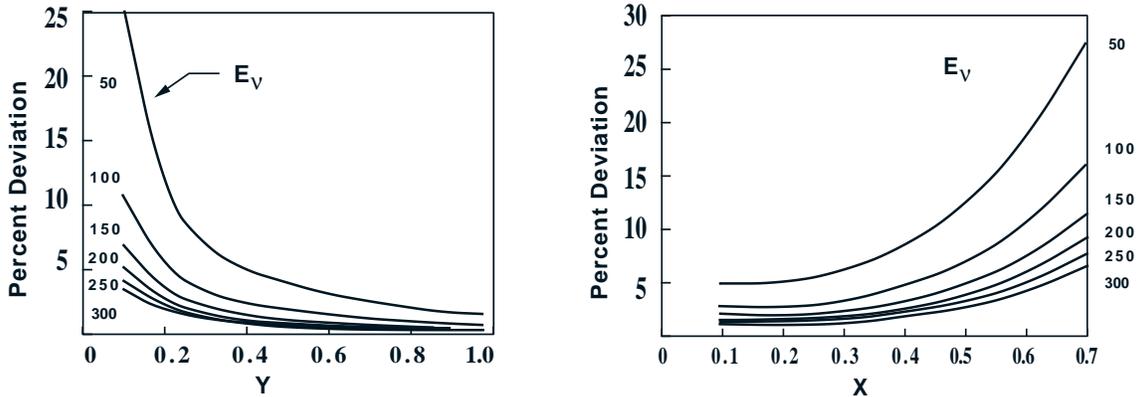}
\caption{ Percent deviation of leading-order cross
section between the ``slow-rescaling,"
Eq.~(\protect\ref{eq:slowresxsii}), and  and complete,
Eq.~(\protect\ref{eq:dsdxynu}), for  $E_\nu=80GeV$, $m_c=1.5GeV$:
(a)  $d\sigma/dy(\nu+s\to c)$  integrated in $x$ over the range 
$x=[0.1,0.6]$; (b)  $d\sigma/dx(\nu+s\to c)$ integrated in $y$ over
the range $y=[0.1,0.8]$}
  \label{fig:SRdy}
\end{figure}
}
\def\xfigbasicproc{
\begin{figure}[hb]
 \epsfxsize=\hsize
  \epsfbox{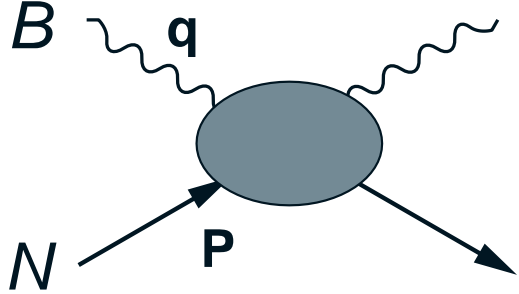}
\caption{Basic process for inclusive boson B(q) nucleon N(P)
scattering: $ N(P) + B(q) \rightarrow X(P_X)$, summed over the final
state, $X(P_X)$} 
  \label{fig:basicproc}
\end{figure}
}
\def\xfigBWi{
\begin{figure}[hb]
 \epsfxsize=\hsize
  \epsfbox{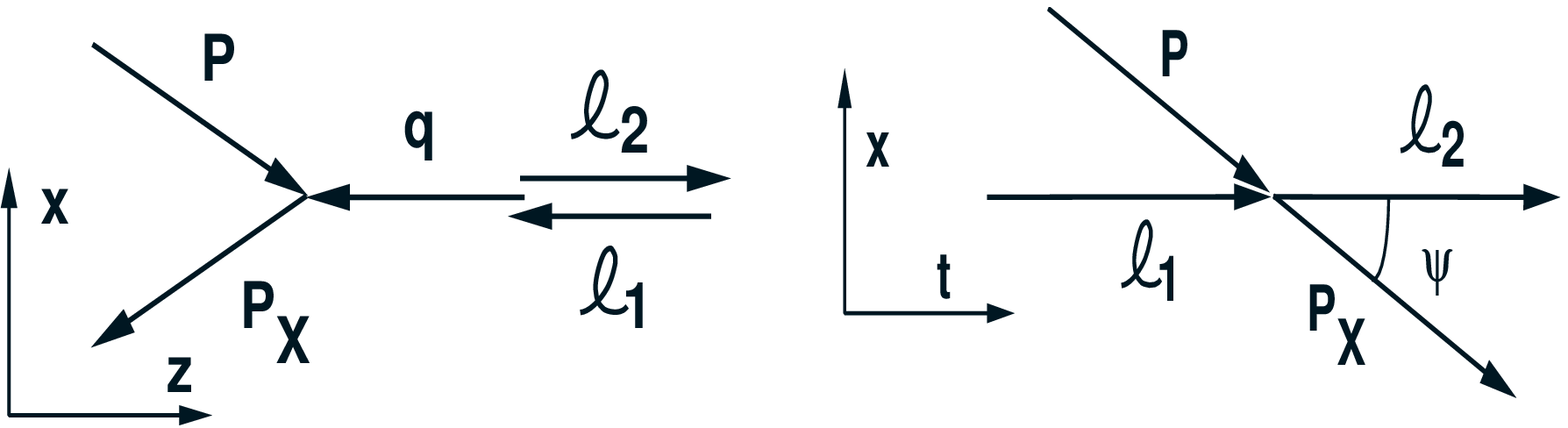}
\caption{(a) The standard lepton configuration in $\{ x,z\}$-space. 
Note that the lepton momenta are colinear with the $z$-axis, and the
hadron momenta define the $x-z$ plane; (b) The same fame seen in $\{ x,t\}$-space.} 
  \label{fig:BW1}
\end{figure}
}
\def\xfigBWii{
\begin{figure}[ht]
 \epsfxsize=\hsize
  \epsfbox{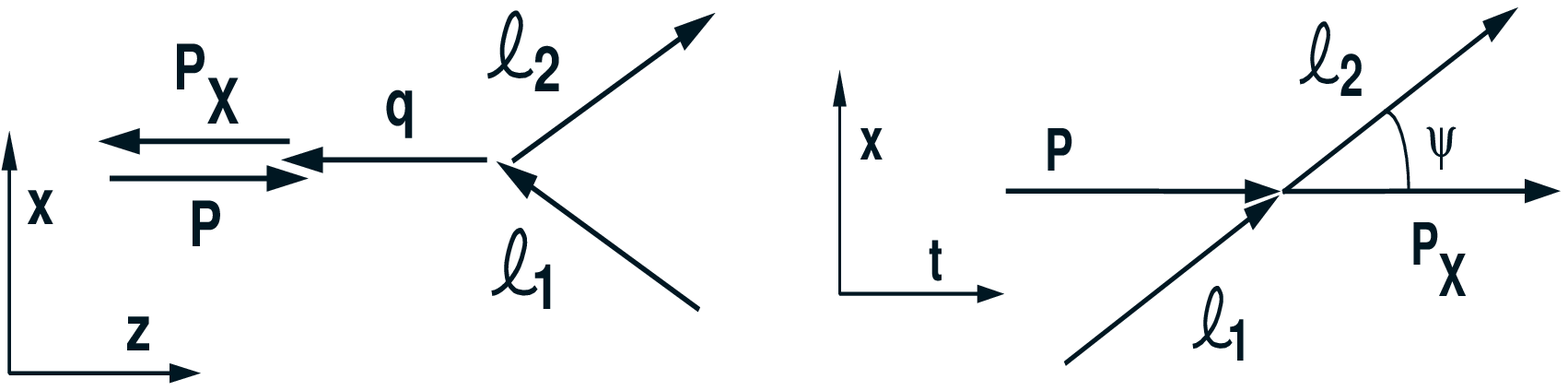}
\caption{ (a) The standard hadron configuration in $\{ x,z\}$-space. 
Note that the hadron momenta are colinear with the $z$-axis, and the
lepton momenta define the $x-z$ plane. (b) This frame
is related to the standard lepton configuration (Fig.~{\protect \ref{fig:BW1}} below)
 by a  space-time
rotation (\ie boost) in the $\{ x,t\}$-plane by the angle $\psi$. 
}  
  \label{fig:BW2}
\end{figure}
}

\epsfverbosetrue


\null
\vfil

\pagestyle{empty}
\begin{center}
\begin{tabular}{l}
October 1993
\end{tabular}
\hfill
\begin{tabular}{l}
hep-ph/9312318  \\
MSU-HEP 93/15           \\
SMU-HEP/93-16           \\
\end{tabular}
\\[1cm]

{\LARGE \ Leptoproduction of Heavy Quarks  I}\\[10pt]

{\Large -- \ General Formalism and Kinematics of Charged Current and 
\\[0.1in]
Neutral Current Production Processes}\\[1cm]

{\large \ M. A. G. Aivazis,${}^a$ 
Fredrick I. Olness,${}^a$\footnote{SSC Fellow}
and Wu-Ki Tung${}^b$}
\\[0.5in]

${}^a$Southern Methodist University, Dallas, Texas 75275 \\

${}^b$Michigan State University, East Lansing, MI 48824
\end{center}
\vfil

\begin{abstract}
Existing calculations of heavy quark production in
charged-current and neutral current lepton-hadron scattering are formulated
differently because of the artificial distinction of ``light'' and ``heavy''
quarks made in the traditional approach. A proper QCD formalism valid for a
wide kinematic range from near threshold to energies much higher then
the quark mass should
treat these processes in a uniform way. We formulate a unified approach to
both types of leptoproduction processes based on the conventional factorization 
theorem. In this paper, we present the general framework with complete
kinematics appropriate for arbitrary masses, emphasizing the simplifications
provided by the helicity formalism. We illustrate this approach with an
explicit calculation of the leading order contribution to the quark
structure functions with general masses. This provides the basis for a
complete QCD analysis of charged current and neutral current leptoproduction
of charm and bottom quarks to be presented in subsequent papers. 
\end{abstract}

\vfil 

\noindent
PACS numbers: 12.38.Bx, 11.10.Gh, 13.60.Hb \\
Journal-ref: Phys.Rev. D50 (1994) 3085-3101

\newpage
\voffset=0.0in 
\pagestyle{plain}




\section{ Introduction \label{sec:Intro}}

Total inclusive lepton-hadron deep inelastic scattering has been the
keystone of the quark-parton picture and the QCD-based Parton Model. As the
global QCD analysis of high energy interactions becomes more precise, other
processes begin to play an increasingly important role in determining the
parton distributions inside the 
nucleon\rlap.\rf{Snowmass,CTEQ1,MT,tungowens} For
instance, semi-inclusive charm production in charged-current and
neutral-current interactions in lepton-hadron scattering serves as a unique
probe of the strange quark and charmed quark content of the nucleon\rlap.%
\rf{CCFR,FMMF,CDHSW} In general, the production of heavy flavors in lepton-hadron
and hadron-hadron colliders is a very important tool for quantitative QCD
study and for searches for new physics\rlap.%
\rf{datareview,Fogli,Halzen,FurPet,bij,lampe,smith,CHARM,EMC}

Traditional analysis of massive quark production in DIS uses the simple
light flavor parton model formulas (based on tree-level forward Compton
scattering off the quark) with a ``charm threshold'' or ``slow-rescaling''
correction\rlap.\rf{slowrescale,Gottschalk,Brock} This prescription is still
widely used in current literature, particularly for dimuon production in
neutrino charged current scattering\rlap.\rf{CCFR,FMMF,CDHSW}; however, the
applicable range of this approach is very limited -- for the neutral current
case by the mass of the initial state quark; and for both cases, by the
numerically important next order gluon contribution.$^{\cite{aot}}$ In most
neutral-current charm production calculations and recent HERA studies of
heavy flavor production, a contrasting view has been provalent: one forsakes
leading-order quark scattering mechanism and concentrates on the ${\cal O}%
(\alpha _s)$ ``gluon-fusion'' processes\rlap.\rf{ali,schuler} Whereas this
latter approach is appropriate when the hard scattering scale of the
process, say $Q$, is of the same order of magnitude as the quark mass $m$\rf{CSS86},
it is a poor approximation at high energies. In fact, when $m/Q$ is small,
these ``gluon-fusion'' diagrams contain large logarithms, \ie factors of the
form $\alpha _s^n\log ^n(m/Q),$ which vitiates the perturbation series as a
good approximation. These large logarithms need to be resummed, which then
yield quark-scattering contributions with properly evolved parton
distribution for the not-so-heavy massive quark. 

A consistent QCD analysis of this problem requires a renormalization scheme
which contains the two conventional approaches as limiting cases---in their
respective region of validity---and provides a smooth transition in the
intermediate region where neither approximation is accurate. Such a scheme,
motivated by the Collins-Wilczek-Zee\rf{cwz} renormalization procedure, was
proposed some time ago in the context of Higgs production\rlap, resulting in
a satisfactory theory valid from threshold to assymptotic energies.
\rf{tungolness} This approach also provides a natural framework for heavy
quark production. It is particularly simple to implement in leptoproduction
production processes, and has been applied to charm production in DIS in a previous
short report\rlap.\rf{aot}

The current paper is the first of a series which will give a
detailed formulation of this problem. In systematically developing a
consistent formulation of heavy-flavor production in DIS, one finds that
conventional calculations, {\em even at the leading order level}, make
implicit approximations inherited from the zero-mass parton model--such as
the Callan-Gross relation and the choice of the scaling variable--which are
not always valid in the presence of masses. 
In order to make a fresh start on a consistent theory including non-zero-mass
partons, this first paper is
devoted to a self-contained development of the general formalism of deeply
inelastic scattering in the presence of masses which is valid for both
charged and neutral current interactions. Much of this is kinematical in nature. 
In considering charm production in
existing fixed target neutrino experiments, an important practical
consideration is that the target nucleon mass is comparable to the charm
quark mass, and {\em both} are non-negligible compared to the average energy
scale $Q$ of the process. Thus, for consistency, target mass effects should
also be incorporated precisely. To this end, we present a helicity formalism
(along with the conventional tensor approach) to develop the general
framework. It will become clear that whereas the conventional tensor method
becomes quite complicated when both target mass and quark mass effects are
properly incorporated, the helicity formalism retains the same simplicity
throughout---due to its group-theory origin, and to a key feature of the QCD
Parton Model. To make the general formalism concrete, we shall apply this helicity
approach to a complete leading order calculation of heavy flavor production
in charged current DIS, and then compare with the conventional tensor
calculations. Numerical studies will show that the complete calculation
(with all masses retained) leads to significant differences in the
calculated cross-sections in certain regions of phase space. In the text of
this paper, we shall emphasize the key elements of these developments. Most
technical details are relegated to the appendices. 

The second paper of this
series\rf{ACOT} shall be focused on the consistent QCD formulation of heavy
quark production in the context of order $\alpha_s$ calculation of this
process, using the general kinematical formalism developed here. The emphasis 
will be on the formulation of a consistent renormalization and factorization
scheme to reconcile the quark-scattering and the gluon-fusion mechanisms.  The 
QCD framework developed there applies to all heavy quark processes, 
including hadroproduction.
In subsequent papers, we shall study the
phenomenological consequences of these calculations on the analysis of
existing dimuon data from fixed target experiments, and on predictions of
charm and bottom production at HERA.


\LaTeXparent{GenForm.tex}

\section{ Scattering Amplitudes \label{sec:ScatAmpl}}

We consider a general lepton-hadron scattering process%
\footnote{
In the production of a heavy quark Q, the final state is given by $X = {\rm Q} +
X'$ where $X'$ is unobserved. For the purposes of the present discussion, we
shall not single out Q from $X$.  }

\begin{equation}
\label{eq:lhProc}\ell _1(\ell _1)+N(P)\longrightarrow \ell _2(\ell _2)+X(P_X)
\end{equation}
as depicted in Fig.~\ref{fig:lhProc} where the exchanged vector boson ($%
\gamma $, W, or Z) will be labelled by $B$ and its momentum by $q$. 

\xfiglhProc

The
lepton-boson and quark-boson couplings are specified by the following
generic expression for the effective fermion-boson term in the electro-weak
Lagrangian: 
\begin{equation}
\label{eq:EwIntL}{\cal L}_{{\rm int}}^{{\rm EW}}=-g_B\left[ j_\mu ^{(\ell
)}(x)+J_\mu ^{(h)}(x)\right] \ V_B^\mu (x)
\end{equation}
where a summation over $B$ is implied. The gauge coupling constant $g_B$ for
the vector boson field $V_B$ depends on B and their values as prescribed by
the Standard Model are given in Table~\ref{tab:gaugeCoupl}.

\begin{table}[tb]
\[
        \begin{array}{c|c|c|c}
            \rule[-.5cm]{0cm}{1.2cm}
            \makebox[.75in][c]{$B$} &
            \makebox[.75in][c]{$\gamma$} &
            \makebox[.75in][c]{$W^{\pm}$} &
            \makebox[.75in][c]{$Z$}
        \\ \hline
            \rule[-.5cm]{0cm}{1.2cm}
            g_{B} &
            - e &
            \displaystyle \frac{g}{2\sqrt{2}} &
            \displaystyle \frac{g}{2\cos \theta_{W}}
        \end{array}
\]
  \caption{The gauge couplings of the vector 
   bosons according to the Standard Model.
   \label{tab:gaugeCoupl}  }
\end{table}

Both the hadronic and fermionic current operators are defined by
\begin{equation}
\begin{array}{ccl}
J_\mu ^{(f)}(x) & = & \overline{\psi }_f(x){\gamma _\mu \ }(g_V-g_A{\gamma
^5\ })\psi _f(x) \\  & = & \overline{\psi }_f(x){\gamma _\mu \
}[g_R(1+{\gamma ^5\ })+g_L(1-{\gamma ^5\ })]\psi _f(x)
\end{array}
\end{equation}
where $\psi _f$ denotes a generic fermion field, and the vector and axial
vector couplings $g_{V,A}$ are related to their chiral counterparts by $%
g_{L,R}$ by $g_{V,A}=g_L\pm g_R$. The values of those fermion coupling
constants, according to the Standard Model, are given in 
Table~\ref{tab:chCoupl}; however, we will keep them general in our considerations.

The scattering amplitude for the process of Eq.~(\ref{eq:lhProc})---with
particle momenta as shown in Fig.~\ref{fig:lhProc}---is given by 
\begin{equation}
\label{eq:InvAmplT}{\cal M} = J_{\mu}^{*} (P,q) \frac{g_B^2 G^{\mu}{}_{\nu} 
}{Q^2 + M_{B}^{2}} j^{\nu} (q,\ell) 
\end{equation}
where $q=\ell_{1}-\ell_{2}$, $\ell = \ell_{1}+\ell_{2}$, $Q^2=-q^2>0$, and $%
G^{\mu}{}_{\nu} = g^{\mu}{}_{\nu} - q^{\mu} q_{\nu} / M_{B}^{2}$. The lepton
current matrix element is given by 
\begin{equation}
\label{eq:lepCurT} j^{\mu} (q,\ell) = \langle \ell_{2} | j^{\mu} | \ell_{1}
\rangle = \overline{u}(\ell_{2}) {\gamma^{\mu} } [ g_{R} (1 + {\gamma^{5} })
+ g_{L} (1 - {\gamma^{5} }) ] u(\ell_{1}) 
\end{equation}
The hadron current matrix element is kept in the general form: $J_{\mu}^{*}
(P,q) = \langle P_{X} | J_\mu^{\dagger} | P \rangle $. For simplicity, we
have suppressed the polarization indices for all external particles in Eq.~(%
\ref{eq:InvAmplT}). Furthermore, the term $G^{\mu}_{\nu}$ can be replaced by 
$g^{\mu}_{\nu}$ in actual applications since the term proportional to $%
q^{\mu}q_{\nu}$ (when contracted with the lepton current matrix element)
yields terms proportional to $m_{\ell}^{2} / Q^{2}$ which are negligible at
high energies.

An alternative expression to the above familiar formulation of the
scattering amplitude which emphasizes the helicity of the exchanged vector
boson is given by:\cite{helForm1,helForm2} 
\begin{equation}
\label{eq:InvAmplH}{\cal M}=J_m^{*}(Q^2,P\negthinspace \cdot \negthinspace
q)\ \frac{g_B^2\ d^1(\psi )^m{}_n}{Q^2+M_B^2}\ j^n(Q^2)
\end{equation}
where $n$ and $m$ are helicity indices for the vector boson, $j^n(Q^2)$ and $%
J_m^{*}(Q^2,P\negthinspace \cdot \negthinspace q)$ are the scalar helicity
amplitudes for the two vertices shown in Fig.~\ref{fig:lhProc}, and $%
d^1(\psi )$ is a spin~1 ``rotation'' matrix specifying the relative
orientation of the two vertices. The derivation of this formula can be found
in$^{\,\cite{helForm1,helForm2}}$; the precise definition of the rotation
angle%
\footnote{
For space-like $q$, $\psi$ is actually a hyperbolic angle specifying a
Lorentz boost.} $\psi $ is given in Appendix~\ref{app:Kin}. (See also
Appendix~\ref{app:SF} for details). We note that the structure of Eq.~(\ref
{eq:InvAmplH}) is quite similar to Eq.~(\ref{eq:InvAmplT}) above. The
advantages of using the helicity formulation in the QCD analysis of heavy
quark production will be discussed in Section~\ref{sec:FacThm}.

\begin{table}[tb]
\[
\begin{array}{|c||c|c|c|}
\hline \ttstrut
 & \gamma & Z & W^\pm \\ \hline 
 \hline \ttstrut
g_V & Q_i & T_{3L}^{i} - 2 Q_i \sin^2\theta_W  
                            & 1 \cdot V_{ij} \\ \hline \ttstrut
g_A & 0   & T_{3L}^{i}             & 1 \cdot V_{ij} \\ \hline 
 \hline \ttstrut
g_R & \frac{Q_i}{2} &  - Q_i \sin^2\theta_W            & 0 \\ \hline \ttstrut
g_L & \frac{Q_i}{2} & T_{3L}^{i} - Q_i \sin^2\theta_W  
              & 1 \cdot V_{ij} \\ \hline
 \hline
\end{array}
\]
  \caption{The gauge couplings of the vector bosons according to the Standard
Model.  $V_{ij}$ represents the CKM flavor mixing, if relevant, and
$Q_i$ is the fermion charge in units of $|e|$. 
  \label{tab:chCoupl}   }
\end{table}


\LaTeXparent{GenForm.tex}

\section{ Cross-section Formulas and Hadron Structure Functions \label
{sec:StrFun}}

The cross-section formula for this process is ({\it cf}. Appendix~\ref{app:Kin}), 
\begin{equation}
\label{eq:dSigma}d\sigma =\frac{G_1\,G_2}{2\Delta (s,m_{\ell _1}^2,M^2)}\
4\pi Q^2\,L_\nu ^\mu \,W_\mu ^\nu \,d\Gamma 
\end{equation}
where $G_i=g_{B_i}^2/(Q^2+M_{B_i}^2)$ is a short-hand for the boson coupling
and propagator. The two indices $B_1$ and $B_2$ denoting the species of the
exchanged vector bosons are implicitly summed over and kept distinct to
accommodate the possibility of $\gamma $-$Z$ interference, and $d\Gamma $ is
the phase space of the final state lepton. The factor $4\pi Q^2$ is from the
normalization of $L$ and $W$. In the above expression we have introduced the
dimensionless lepton and hadron tensors given by%
\footnote{
Historically, the definition of $W^{\mu}{}_{\nu}$---and thus the
definitions of $W_{i}$ in \protect\Eq{invStrFun}---contains an extra factor of
$M$, the target mass. In view of scaling considerations,
it is  more natural to use the
dimensionless definition.  Also note that sums and integrals over all the
unobserved hadronic final states $X$ are implied in
\protect\Eq{hadronTen-Def}.   } 
\begin{eqnarray}
    L^{\mu}{}_{\nu}
        & = &
        \frac{1}{Q^{2}} \
        \overline{\sum_{{\rm spin}}}
            \langle \ell_{1} | j_{\nu}^{\dagger} | \ell_{2} \rangle
            \langle \ell_{2} | j^{\mu} | \ell_{1} \rangle
                                                      \label{eq:leptonTen-Def}
\\
    W^{\mu}{}_{\nu}
        & = &
        \frac{1}{4 \pi}
        \overline{\sum_{{\rm spin}}}
            (2\pi)^{4} {\delta^{4}\!\left(P + q - P_{X}\right)}
            \langle P | J^{\mu} | P_{X} \rangle
            \langle P_{X} | J_{\nu}^{\dagger} | P \rangle
                                                      \label{eq:hadronTen-Def}
\end{eqnarray}
The explicit expression for $L^\mu {}_\nu $ with general coupling constants
is given in Appendix~\ref{app:SF}. As is well known, the hadron tensor $%
W^\mu {}_\nu $ can be expanded in terms of a set of six independent basis
tensors%
\footnote{
In some papers, the tensor associated with $W_1$ is chosen to be the gauge
invariant form $(-g^\mu_\nu + q^\mu q_\nu /q^2)$,
and that associated with $W_2$ is obtained with the  substitution
$P^{\mu}~\rightarrow~P^{\nu}(g^\mu_\nu - q^\mu q_\nu /q^2)$;
these changes (convenient for conserved currents) will modify
the definitions of $W_{4}$, $W_{5}$ and $W_{6}$ only.} 
\begin{eqnarray}
    W^{\mu}{}_{\nu}
        & = &
      - g^{\mu}{}_{\nu} W_{1}
      + \frac{P^{\mu} P_{\nu}}{M^2}  W_{2}
      - i \frac{\epsilon^{Pq\mu}{}_{\nu}}{2 M^2} W_{3} +
\label{eq:invStrFun}
\\ & &
      + \frac{q^{\mu} q_{\nu}}{M^2} W_{4}
      + \frac{P^{\mu} q_{\nu} + q^{\mu} P_{\nu}}{2 M^2} W_{5}
      + \frac{P^{\mu} q_{\nu} - q^{\mu} P_{\nu}}{2 M^2} W_{6}
\nonumber
\end{eqnarray}
where $M$ is the target mass and $\epsilon ^{Pq\mu \nu }=\epsilon ^{\alpha
\beta \mu \nu }P_\alpha q_\beta $. The scalar coefficient functions $\{W_i\}$
are the {\em invariant hadron structure functions} for this process.

By substituting the lepton and hadron tensors in Eq.~(\ref{eq:dSigma}) and
partially integrating over the phase space of the final state lepton one
obtains, in the limit of negligible lepton masses, the well-known cross
section formula, generalized to arbitrary couplings, 
\begin{eqnarray}
    \frac{d\sigma}{dE_{2} \, d\cos \theta}
        & = &
        \frac{2 E_2^2}{\pi M}
        \frac{G_1 \, G_2}{n_{\ell}}
     \left\{
            g_{+\, \ell}^{2}
            \left[
                2 W_{1} \sin^2 \frac{\theta}{2} + W_{2} \cos^2 \frac{\theta}{2}
            \right]
          \pm g_{-\, \ell}^{2}
            \left[
                \frac{E_{1}+E_{2}}{M}  W_{3} \sin^2 \frac{\theta}{2}
            \right]
     \right\}
\nonumber \\
 \label{eq:dSigma/dEdz}
\end{eqnarray}
where the $\pm $ sign for the $W_3$ term refers to the case of
lepton/anti-lepton scattering, respectively. Here, $E_1$ and $E_2$ are the
energies of the initial and final state leptons respectively in the
laboratory frame, $\theta $ is the scattering angle of the lepton in the
same frame, and $n_\ell $ is the number of polarization states of the
incoming lepton. To simplify the expression, we define $g_{\pm \,\ell
}^2=g_{L\,\ell }^2\pm g_{R\,\ell }^2$, where $g_{L\,\ell }$ and $g_{R\,\ell }
$ refer to the chiral couplings of the vector boson to the leptons\rlap.%
\footnote{
 The lepton chiral couplings appear explicitly because $L^\mu_\nu$ has been
evaluated.  The corresponding hadron chiral couplings reside in the $\{ W_i \}$
invariant structure functions. }

It is worth noting that the hadron structure functions $\{W_4, W_5, W_6 \}$
do not appear on the right-hand side because they are multiplied by factors
of lepton mass from the lepton vertex, {\em not because they are
intrinsically small} compared to the familiar $\{W_1, W_2, W_3 \}$. This
will become relevant when we discuss the calculation of hard scattering
cross-sections involving heavy quarks.

It is by now customary to introduce the scaling structure functions $F_i$
given by 
\begin{eqnarray}
    F_{1} & = & W_{1}  \nonumber \\
    F_{2} & = & \frac{\nu}{M} W_{2}                   \label{eq:strFun:Fi} \\
    F_{3} & = & \frac{\nu}{M} W_{3} \nonumber
\end{eqnarray}
in terms of which the expression for the differential cross section may be
rewritten as 
\begin{eqnarray}
  \frac{d\sigma}{dx dy}
     & = &
\frac{2M E_1  }{  \pi } \
    \frac{G_1 \, G_2}{n_{\ell}}
  \left\{
        g_{+\, \ell}^2
        \left[
x F_1 \, y^2
+ F_2 \, \left[ (1-y) -  \left( \frac{M x y }{  2 E_1} \right) \right]
        \right]
        \pm  g_{-\, \ell}^2
        \left[
x F_3 \, y ( 1-y/2)
        \right]
    \right\}
\nonumber \\
 \label{eq:dSigma-Fi}
\end{eqnarray}
In the alternative helicity formalism, the expression for the cross section
is given by 
\begin{eqnarray}
  \frac{d\sigma}{dx dy}
     & = &
    \frac{y Q^2}{2 \pi} \,
    \frac{G_1 \, G_2}{n_{\ell}} \,
    \left\{
        g_{+\, \ell}^2
        \left[
            \frac{ (F_{+} + F_{-}) }{2} (1+\cosh^{2}\psi)
        + F_0            \sinh^{2}\psi
        \right]
        \mp g_{-\, \ell}^2
        \left[
            (F_{+} - F_{-}) \cosh\psi
        \right]
    \right\}
\nonumber \\
 \label{eq:dSigma-Hel}
\end{eqnarray}
where $\psi $ is the hyperbolic rotation angle of Eq.~(\ref{eq:InvAmplH}),
and we have introduced the {\em helicity structure functions} $\{F_\lambda $%
, $\lambda =+,0,-\}$ which correspond to the physical forward Compton
scattering helicity amplitudes 
\begin{equation}
\label{eq:StrFun-Hel}F_\lambda =\epsilon _\mu ^\lambda {}^{*}(P,q)\,W^\mu
{}_\nu (P,q)\,\epsilon _\lambda ^\nu (P,q)\makebox[.5in]{}({\rm no\ sum\
over\ }\lambda )
\end{equation}
with {\em right-handed (+)}, {\em longitudinal (0)}, and {\em left-handed (-)%
} vector bosons respectively.%
\footnote{
The choice of these labels---over the more obvious $R$, $L$, etc.---is constrained
by the conflict between the {\em L}eft-handed and {\em L}ongitudinal
designations.
 For $m_\ell=0$, we can ignore $F_\lambda = \{ F_{qq}, F_{q0}, F_{0q} \}$,
{\it cf.}, \App{SF}.   } We note that the first term on the right hand side
involves the {\em transverse} structure function $F_T=(F_{+}+F_{-})/2$,
whereas the third term is the parity-violating term with $F_{+}-F_{-}$
proportional to $F_3$ in Eq.~(\ref{eq:dSigma-Fi}). Eq.~(\ref{eq:dSigma-Hel})
should be familiar, as it is analogous to the corresponding well-known
formul\ae \ for time-like vector boson production processes---Drell-Yan
pairs and W-, Z-production---where the hyperbolic angle $\psi $ is replaced
by the center-of-mass angle $\theta $ for the final state lepton pair.

The helicity structure functions as defined above are naturally scaling
functions. In addition, their direct physical interpretation leads to simple
properties in the QCD parton model framework, as we shall see in the next
section. Note that Eq.~(\ref{eq:dSigma-Hel}) does not show any explicit
target mass dependence; all complications arising from the non-vanishing
mass are contained in the definition of the rotation angle $\psi $ through
kinematics. This simplicity is a consequence of the underlying
group-theoretical approach to the factorized structure of 
Fig.~\ref{fig:lhProc}. The precise relations between the helicity structure functions
and the invariant structure functions are found ({\it cf.}, 
Appendix~\ref{app:SF}) to be: 
\begin{eqnarray}
\begin{array}{rcrcl}
F_+  &=&  F_1  &-& \frac{1}{ 2  }\ \sqrt{1+ \frac{Q^2}{\nu^2} }\  F_3
  \\ [5pt]
F_-  &=&  F_1  &+& \frac{1 }{ 2  }\ \sqrt{1+ \frac{Q^2}{\nu^2} }\  F_3
  \\ [5pt]
F_0  &=& -F_1  &+& \left( 1 + \frac{Q^2 }{ \nu^2 } \right)
           \left( \frac{1 }{ 2x } \right) \  F_2
\end{array}
  \label{FvsW}
\end{eqnarray}
We see in the limit $M\rightarrow 0$ that $Q^2/\nu ^2\rightarrow 0$ and we
obtain the approximation: $F_{\pm }\simeq F_1\mp F_3/2$ and $F_0\simeq
-F_1+F_2/2x$.

To leading order in the electroweak coupling, Eq.~(\ref{eq:dSigma/dEdz}),
Eq.~(\ref{eq:dSigma-Fi}) and Eq.~(\ref{eq:dSigma-Hel}) are completely
general, assuming only Lorentz kinematics and small lepton masses. In
particular, all results up to this point are independent of strong
interaction dynamics. Aside from Eq.~(\ref{eq:dSigma-Hel}), they are
well-established formulae  explicitly generalized to include arbitrary
couplings.


\LaTeXparent{GenForm.tex}

\section{ The QCD Factorization Formulas \label{sec:FacThm}}

Perturbative QCD allows one to relate the measurable hadron structure
functions $\{ F_i \}$ to the corresponding quantities involving elementary
particles---the partons---which can be calculated in perturbation theory.
This section states the basic QCD ``factorization theorem'' as it applies to
deeply inelastic scattering processes and points out some important
unfamiliar features in the presence of non-zero masses, especially when the
initial state parton is a heavy quark.

\subsection{Factorization of Tensor Amplitudes}

The factorization theorem\rf{CSS} states that, in the Bjorken limit, the
dominant contributions to the hadronic tensor structure function has the
factorized form of Fig.~\ref{fig:FactThm} with on-shell, collinear partons: 
\begin{eqnarray}
    W_{\mu \nu}^{B N} (q,P, ...)
        & = &
        \sum_{a}  f^{a}_{N} \otimes \omega_{\mu \nu}^{Ba}
\nonumber \\
        & = &
        \sum_a
        \int {d\xi \over \xi } \
            f^{a}_{N} (\xi, \mu) \
           \omega_{\mu \nu}^{Ba} (q, k_1, ... , \alphas (\mu) )
      \label{eq:FactThm:Wmunu}
\end{eqnarray}

\xfigFactThm

In Eq.~(\ref{eq:FactThm:Wmunu}), the label `$a$' is summed over all parton
species. The convolution integral variable $\xi $ is the momentum fraction
carried by the parton with respect to the hadron defined in terms of the
ratio of light-cone momentum components $\xi =k_1^{+}/P^{+}$. The universal
parton distribution functions $f_N^a$ are scalars; scattering of the vector
boson takes place with the partons via the hard-scattering factor $\omega
_{\mu \nu }^{Ba}$ which can be aptly called the {\em parton structure
function} tensor since it is entirely analogous to the hadron structure
function tensor $W_{\mu \nu }^{BN}$ by substituting the hadron target `$N$'
by the parton target `$a$\rlap.' Note, the tensor structure of $W_{\mu \nu
}^{BN}$ is completely determined by that of $\omega _{\mu \nu }$. These
features should be obvious by inspection of Fig.~\ref{fig:FactThm}. Strictly
speaking, the factorization theorem is established in this simple form only
for certain specifically defined asymptotic regimes. We shall treat 
Eq.~(\ref{eq:FactThm:Wmunu}) as an ansatz and apply it in such a way that our results
reduce to the known correct expressions in the limits $\Lambda \ll m_2\simeq
Q$ on the one hand, and $\Lambda <m_2\ll Q$ on the other.

The presence of heavy quarks among the initial and final state partons in $%
\omega _{\mu \nu }$ has some important consequences. The most immediate one
is that the range of integration in Eq.~(\ref{eq:FactThm:Wmunu}) will depend
on the masses of the heavy quark as a simple consequence of the kinematics
of the hard scattering. In leading order QCD, where the integration range
reduces to a single point, this naturally gives rise to a generalized 
``slow-rescaling'' variable which was originally proposed in the context of
the simple parton model\rlap.\rf{slowrescale} (Cf., Appendix~\ref{app:Kin}.)
In addition, the tensor structure of the perturbatively calculable $\omega
^{\mu \nu }$ is clearly different from that of the naive parton model, even
in leading order QCD! For example, the well-known Callan-Gross relation
simply does not hold in the presence of heavy quark mass. A proper treatment
of heavy quark production must use the correct hard-scattering amplitude $%
\omega _{\mu \nu }^{Ba}$ (calculated to the appropriate order, including
quark masses) in conjunction with choosing the proper variable. A
``slow-rescaling prescription'' of a simple variable substition is not
sufficient, {\it cf.}, Section~\ref{sec:compare}.

In order to apply the factorization theorem to measurable quantities
properly, we must re-express Eq.~(\ref{eq:FactThm:Wmunu}) in terms of the
independent invariant structure functions $\{ W_{i} \}$ or the helicity
structure functions $\{ F_\lambda \}$ in a precise way. Theoretical
calculations of the parton-level hard amplitudes on the right-hand side of
the equation usually yield the (parton) invariant or helicity amplitudes,
not the tensor $\omega^{\mu\nu}$ itself. In the presence of target and heavy
quark masses, we will find that the relations between the invariant
structure functions at the hadron and the parton levels are far from being
simple, as usually assumed in existing literature. In contrast, the
connection between the corresponding helicity structure functions are
completely transparent.

\subsection{Invariant Structure Functions:}

The parton-level invariant amplitudes $\omega_{i}$ are defined in analogy to
Eq.~(\ref{eq:invStrFun}), as follows:%
\footnote{
In order to render the $\omega_{i}$ dimensionless, we use the natural variable
$Q$ rather than any parton mass in scaling the tensors so that the invariant
structure functions have well defined limits as $m/Q \rightarrow 0$.
 (Note, if the hadronic structure functions were originally defined this way,
rather than using the target mass $M$ as the scale factor, $\{W_i\}$ would be
naturally ``scaling!")
} 
\begin{eqnarray}
    \omega^{\mu}{}_{\nu}
        & = &
      - g^{\mu}{}_{\nu} \  \omega_{1}
      + \frac{k_1^{\mu} k_{1 \,\nu}}{Q^2}  \omega_{2}
      - i \frac{\epsilon^{k_1   q\mu}{}_{\nu}}{2 Q^2} \omega_{3} +
\nonumber \\ & & \makebox[.15in]{}
      + \frac{q^{\mu} q_{\nu}}{Q^2} \omega_{4}
      + \frac{k_1^{\mu} q_{\nu} + q^{\mu} k_{1\,\nu}}{2 Q^2} \omega_{5}
      + \frac{k_1^{\mu} q_{\nu} - q^{\mu} k_{1\,\nu}}{2 Q^2} \omega_{6}
                                                  \label{eq:PrtnStrFun}
\end{eqnarray}
where $k_1$ is the momentum of the incident parton. 
Substituting Eq.~(\ref{eq:PrtnStrFun}) in Eq.~(\ref{eq:FactThm:Wmunu}) 
and comparing 
$\omega^{\mu}{}_{\nu}$ with $W^{\mu}{}_{\nu}$ (Eq.~(\ref{eq:invStrFun})), we
see that the relations between invariant structure functions at the hadron
and the parton levels depend on the relation between $k_1^\mu$ and $P^\mu$.
Whereas the two momenta are proportional in the zero mass limit, this
relation becomes non-trivial in the presence of {\it either} target mass or
parton mass, ({\it cf.}, Appendix~\ref{app:Kin}). Since the vectors $P$, $k_1
$ and $q$ are collinear, we can parametrize $k_1$ as 
\begin{equation}
k_1^{\mu} = \zeta_{P} P^{\mu} + \zeta_{q} q^{\mu} 
\end{equation}
In the zero mass limit, $\zeta_{P} \rightarrow \xi$ and $\zeta_{q}
\rightarrow 0$. In general, the coefficients ($\zeta_{P}, \zeta_{q}$) are
rather complicated functions of the masses and the {\em convolution variable}
$\xi$, ({\it cf.}, Eq.~(\ref{eq:kqPrel})). Thus, the relations between the $%
W_{i}$ and the $\omega_{i}$ are also rather complicated. Relevant formulas
which relate $W_{i}$ to $\omega_{i}$ are given in Appendix~\ref{app:SF}.

\subsection{Helicity Structure Functions:}

In sharp contrast to the above, the factorization theorem assumes a simple
form when expressed in terms of the helicity basis. To see this, let us
define the parton {\em helicity structure functions} $\omega_{\lambda}$, in
analogy to Eq.~(\ref{eq:StrFun-Hel}), by: 
\begin{equation}
\label{eq:PtnStrFun-Hel} \omega_{\lambda} =
\epsilon^{\lambda}_{\mu}{}^{*}(k_1,q) \: \omega^{\mu}{}_{\nu} \:
\epsilon_{\lambda}^{\nu}(k_1,q) \makebox[.5in]{} ({\rm no\ sum\ over\ }%
\lambda) 
\end{equation}
In order to relate these to the hadron helicity structure functions $%
F_{\lambda}$, Eq.~(\ref{eq:StrFun-Hel}), it appears that one needs to
re-express the vector-boson polarization vectors \{$\epsilon_{\lambda}^{%
\nu}(k_1,q)$\} (defined using $k_1$ as the reference momentum) in terms of \{%
$\epsilon_{\lambda}^{\nu}(P,q)$\} (defined using $P$ as the reference
momentum). The enormous simplification of the helicity approach follows from
the fact that {\em the two sets of polarization vectors are in fact
identical even in the presence of masses}, hence no transformation is
needed! The reason for this is that for a given vector-boson momentum $q$,
the reference momentum is used only to specify the direction of the
polarization axis; the two seemingly different reference momenta $k_1$ and $%
P $ actually specify the same set of polarization vectors because they are
collinear in the QCD Parton framework. Thus, we arrive at the
straightforward formula: 
\begin{equation}
\label{eq:FactThm-Hel} F_{\lambda}^{B N} (q,P, ...) = \sum_{a} f^{a}_{N}
\otimes \omega_{\lambda}^{B,a} 
\end{equation}

This suggests that to explore the consequences of perturbative QCD on heavy
quark production (as well as on all other processes), it is advantagous to
perform the calculation in the helicity basis. The simple formula 
Eq.~(\ref{eq:FactThm-Hel}), together with Eq.~(\ref{eq:dSigma-Hel}), relate the
calculation of hard scattering amplitudes directly to measurable
cross-sections without any approximations or complications. Besides, since
the parton-level helicity amplitudes have simple symmetry and structure, due
to the basic chiral couplings of the theory, the results of this approach
are often the most physical and compact to begin with.


\LaTeXparent{GenForm.tex}

\section{ Leading Order QCD Calculation of Heavy Flavor Production \label
{sec:LO}}

To illustrate the use of the general formalism developed above, we apply it
to the calculation of heavy quark production in leading order QCD. Existing
applications of heavy quark production in DIS mostly concern charm
production in charged current interactions at fixed-target energies. Since
the charm mass is comparible to the target mass for existing neutrino
experiments, and neither is negligible compared to the energy scale $Q$, it
is reasonable to retain the target mass effects in order to be self
consistent. Numerical comparisons of the complete calculation (with full
target mass dependence) to the conventional one show that the difference can
be significant in certain regions of the phase space.

\xfigBorn

The leading order diagram that contributes to $\omega _\lambda $ is shown in
Fig.~\ref{fig:Born} and its contribution, including all masses and arbitrary
couplings, is calculated explicitly in Appendix~\ref{app:LO}. We consider
charm production in charged current neutrino scattering. Since, the $W$%
-exchange process involves only left-handed chiral couplings, ({\it cf.},
Table~\ref{tab:chCoupl}). The parton helicity structure functions for
scattering from a strange quark are given by 
\begin{eqnarray}
    \omega_{\pm}
        & = &
        g_{L\, a}^{2} \:
        \frac{Q^{2}+m_{1}^{2}+m_{2}^{2} \mp \Delta}{\Delta} \:
        {\delta\!\left(\frac{\xi}{\chi} - 1\right)}
\nonumber \\
    \omega_{0}
        & = &
        g_{L\, a}^{2} \:
        \frac{(m_{2}^{2}-m_{1}^{2})^{2} / Q^{2} + m_{2}^{2} + m_{1}^{2}}{\Delta}
        \:{\delta\!\left(\frac{\xi}{\chi} - 1\right)}
\label{eq:LO-omg}
\end{eqnarray}
where $g_{L\, a}^2$ is the left-handed coupling of the $W$ to the $a$-type
parton, $\xi $ is the convolution variable of Eq.~(\ref{eq:FactThm:Wmunu}), $%
m_1$ is the initial parton mass, $m_2$ is the heavy quark mass, and $\chi $
and $\Delta $ are given by 
\begin{eqnarray}
    \chi & = & \eta \
{ (Q^2-m_1^2+m_2^2) + \Delta   \over 2 Q^2 }
\\
    \Delta & = &  \Delta[-Q^2,m_1^2,m_2^2]
\end{eqnarray}
where $\eta $ (Eq.~(\ref{eq:etagen})) is the target-mass corrected Bjorken $x
$, and $\Delta $ is the triangle function (Eq.~(\ref{eq:triangle})), both
defined in Appendix~\ref{app:Kin}.

Substituting in Eq.~(\ref{eq:FactThm-Hel}), we obtain simple but non-trivial
formulas for the hadron helicity structure functions. The delta function in
Eq.~(\ref{eq:LO-omg}) fixes the momentum fraction variable $\xi =\chi $.
Since $\omega _0\not =0$, we see explicitly that the longitudinal structure
function cannot be neglected {\it even to leading order}. It is proportional
to the quark masses when they are non-vanishing; thus, the Callan-Gross
relation does not apply in its original form.

For charm-production, the initial parton is either a $d$ or $s$ quark; both
can be treated as massless. In the limit $m_1\rightarrow 0$, one obtains%
\begin{eqnarray}
    \omega_{+} & = &  0 \\
    \omega_{-} & = & g^{2}_{L\, a} \, 2 \, \delta(\xi/\chi-1) \\
    \omega_{0} & = & g^{2}_{L\, a} \frac{m_2^2}{2 Q^2}  \, 2 \, \delta(\xi/\chi-1)
\end{eqnarray}
and $\chi =\eta (1+m_2^2/Q^2)$. Thus, the helicity structure functions
assume the following simple form:%
\begin{eqnarray}
    F_{+} & = &  0 \\
    F_{-} & = & g^{2}_{L\, a} \, 2 \, q^{a}_{N}(\chi) \\
    F_{0} & = & g^{2}_{L\, a} \frac{m_2^2}{2 Q^2}  \, 2 \, q^{a}_{N}(\chi)
\end{eqnarray}
where an implicit sum over contributing parton species $a$ is implied. By
applying the general expression of Eq.~(\ref{eq:dSigma}), one obtains%
\begin{eqnarray}
    \frac{d\sigma^{\nu}}{dx dy}
        & = &
          G_W^2 \, g_{L\, \ell}^2  g_{L\, a}^2 \, 2 \, q^{a}_{N}(\chi) \
        \frac{y Q^2}{\pi}
        \left[
            \left( \frac{1+\cosh\psi}{2} \right)^2
          + \frac{m_{2}^{2}}{2 Q^{2}} \frac{\sinh^{2}\psi}{2}
        \right]
	\label{eq:dsdxynu}
\end{eqnarray}
where $\psi $ is defined by Eq.~(\ref{eq:coshPsi}), $g_{L\,\ell }=1$ and 
$g_{L\, a}=\cos \theta _C(\sin \theta _C)$ for $a=s(d)$, respectively. Note, 
$G_W=g_{B_W}^2/(Q^2+M_W^2)=(G_F/\sqrt{2})/(1+Q^2/M_W^2)$.

The corresponding formula for anti-quark production via lepton scattering,
obtained from the interchange of $g_{L\, a}$ and $g_{R\, a}$ in the expressions
for $\omega_\lambda$, yields: 
\begin{eqnarray}
    F_{+} & = & g^{2}_{L\, {\overline{a}}} \, 2 \,
                            \overline{q}^{\overline{a}}_{N}(\chi) \\
    F_{-} & = &  0 \\
    F_{0} & = & g^{2}_{L\, {\overline{a}}} \frac{m_2^2}{2 Q^2}  \, 2 \,
                            \overline{q}^{\overline{a}}_{N}(\chi)
\end{eqnarray}
and 
\begin{eqnarray}
    \frac{d\sigma^{\overline{\nu}}}{dx dy}
        & = &
          G_W^2 \, g_{L\, \ell}^2  g_{L\, {\overline{a}}}^2 \, 2 \,
                     \overline{q}^{\overline{a}}_{N}(\chi)\
        \frac{y Q^2}{\pi}
        \left[
            \left( \frac{1-\cosh\psi}{2} \right)^2
          + \frac{m_{2}^{2}}{2 Q^{2}} \frac{\sinh^{2}\psi}{2}
        \right]
	\label{eq:dsdxynubar}
\end{eqnarray}

These results still retain the full kinematic target-mass dependence ({\it 
cf.}, Appendix~\ref{app:Kin}). If one sets $M=0$, the expressions for the
cross section in Eqs.~(\ref{eq:dsdxynu}) and (\ref{eq:dsdxynubar}) stay
unchanged; only the definitions of $\psi $ and $\chi $ simplify. In
particular
\begin{equation}
\chi \stackunder{m_1\rightarrow 0}{\longrightarrow }\eta 
\left( 1+\frac{m_2^2}{Q^2}\right) 
\stackunder{M\rightarrow 0}{\longrightarrow }x
\left( 1+\frac{m_2^2}{Q^2}\right) 
\end{equation}
which is the ``slow-rescaling'' variable.


\LaTeXparent{GenForm.tex}

\section{ Comparison with Existing Calculations \label{sec:compare}}

There are a variety of ``slow-rescaling'' prescriptions in the literature
with varying degrees of accuracy.\rf{slowrescale} Some analyses of charm
production in DIS use a slow-rescaling corrected parton model prescription
which consists of using the familiar zero-mass parton model cross-section
with the substitution: 
\begin{eqnarray}
x \rightarrow \xi  = x   \left( 1 + \frac{m_2^2}{ Q^2} \right)
\end{eqnarray}
This prescription incorporates only the heavy quark mass effect for the
on-mass shell kinematics---the delta function of Eq.~(\ref{eq:LO-omg})---but
ignores corrections to the ``body'' of the partonic (hard) structure
functions $\omega _\lambda $ in the same equation. It is therefore
inherently inconsistent.

An improved treatment is obtained by using the exact expression for the Born
diagram with $m_1=0$ and $M=0$. The results are simple enough so that the
final $m_2$ dependence can be rewritten to appear as a ``slow-rescaling''
corrected formula, as follows: 
\begin{eqnarray}
    \frac{d\sigma}{dx dy}
        & = &
         G_W^2  \, g_{L\, \ell}^2  g_{L_a}^2  \
        \frac{ 2 Q^2}{ \pi y }\
\left\{
\left[ y + \frac{\xi }{ x }\ (1-y)   \right]
q(\xi)  +
\left[ y(y-1) + \frac{\xi}{x }\ (1-y)   \right]
\bar{q}(\xi)
\right\}
\nonumber \\
\label{eq:slowresxs}
\end{eqnarray}
By definition, this modified prescription ignores target masss effects in
the parton kinematics that are not necessarily small compared with heavy
quark effects. Eq.~(\ref{eq:slowresxs}) should be compared with 
Eq.~(\ref{eq:dsdxynu}) which has implicit $M$ dependence in 
$\cosh\psi$, $\sinh\psi$, and $\chi$.

Some papers include the target mass dependence of the cross section  Eq.~(%
\ref{eq:dSigma-Fi}), i.e., the term $-Mxy/(2E_1)$, so that the cross section
for neutrino production reads: 
\begin{eqnarray}
    \frac{d\sigma^{\nu}}{dx dy}
        & = &
         G_W^2  \, g_{L\, \ell}^2  g_{L_a}^2  \
        \frac{ 2 Q^2}{ \pi y }\
\left\{  y + \frac{\xi}{x }\ (1-y)
   - \frac{\xi}{x }\ \left(\frac{M x y}{2 E_1}  \right)
\right\} \ q(\xi)
\nonumber \\
\label{eq:slowresxsii}
\end{eqnarray}
Numerically, this term has negligible effect; the $-Mxy/(2E_1)$ term does
not approximate the true target mass dependence, and for all practical
purposes, Eq.~(\ref{eq:slowresxs}) and Eq.~(\ref{eq:slowresxsii}) are
identical at the $\leq 2\%$ level.

\xfigSRdy

We now present numerical results comparing cross-sections calculated using
the complete leading order formula Eq.~(\ref{eq:dsdxynu}) with that using
the slow-rescaling prescription, Eq.~(\ref{eq:slowresxsii}). 
In Fig.~\ref{fig:SRdy}we 
compare the $y$ and $x$ dependence for $\nu
+N\rightarrow \mu ^{-}+c+X$ for neutrino energies ranging from 50\thinspace
GeV to 300\thinspace Gev---a reasonable range for fixed target experiments.
For simplicity, we only consider the dominant sub-process: $W+s\rightarrow c$
. As anticipated, for both the $x$ and $y$ distributions, the deviations
decrease with increasing neutrino energy, (hence, increasing $Q^2$) since
the $M^2/Q^2$ and $m_2/Q^2$ terms are decreasing. The $y$ distribution
agrees well at large $y$, but deviates from the complete leading-order
result by more than 25\% for small $y$ where the effects of the charm mass
threshold are significant. The deviation of the $x$ distribution ranges from
a few percent at small $x$ to $\geq 25\%$ at large $x$. Thus the difference
between the conventional slow-rescaling prescription and our approach, which
is based on the factorization theorem, are not negligible. The main source
of discrepancy arises from the charmed quark mass $m_2$ which is only
slightly larger than the target mass $M$; the latter should not be neglected
if effects due to the former are significant. In particular, the momentum
fraction variable $\xi =\chi $ which enters the precise formula 
Eq.~(\ref{eq:dsdxynu}) is approximately: 
\begin{eqnarray} \xi  &=&
\chi \quad  \simeq  \quad
 x_ {} \
\left( 1 + \frac{m_2^2}{Q^2}  \right) \
\left( 1 - \frac{x ^2 M^2 }{ Q^2}  \right)
\end{eqnarray}
when $m_2^2/Q^2$ and $M^2/Q^2$ are small, and $m_1=0$. In other words, the
conventional ``slow-rescaling'' variable itself needs a target-mass
correction.


\LaTeXparent{GenForm.tex}

\section{ Conclusions \label{sec:conclusions}}

The proper treatment of the effects of heavy quarks in the theoretical
predictions of the differential cross section for deeply inelastic
scattering processes is not completely solved in perturbative QCD. Strictly
speaking, the familiar factorization theorem applies only to one scale
problems, {\it i.e.}, when either all quark masses are negligible compared to $Q^2$,
or when the heavy quark mass $m$ is of the same order of magnitude as $Q^2$.

The recent higher order calculations of heavy quark production which exclude
massive partons and focus on the gluon-fusion diagrams apply only to the
region in which $m^2\sim Q^2$ and require a totally different treatment of
charged and neutral  current processes.

We formulate a unified approach to both types of processes that is based on
the factorization theorem as an ansatz. We assume that the factorization
theorem holds throughout the energy range of interest in the simple form $%
W=f\otimes \omega $. This ansatz produces the correct results in the regimes
$Q^2\sim m^2$ and $Q^2\gg m^2$, and provides a {\em smooth} interpolation in
the intermediate regions. We are able to treat both charged and neutral
current processes by endowing the parton quarks with a mass and by not
making {\em a priori }any assumptions about the relative importance of quark
and gluon-initiated contributions. Instead, we take advantage of precisely
the techniques that yield the proof of the factorization theorem to ensure
that the final expressions conform to expectations in the
$Q^2\sim m^2$ and $Q^2\gg m^2$ regions.

Working towards this goal, we have presented here the general framework. In
order to illustrate the basics of our approach, we have presented an explicit
calculation of the lowest order contribution to the quark structure
functions. However, this contribution by itself is not sufficient for proper
phenomenological analysis of DIS cross sections because of the importance of
quark-gluon mixing in sea-quark initiated processes.

We have compared  existing phenomenological analyses based on the lowest order
process $W+q \rightarrow Q$, with the unified approach which retains all masses.
For charged current  charm production experiments ($W+s \rightarrow c$), the final
state heavy quark mass $m$ is comparable to the target mass $M$; hence, if the
$m$-dependence is retained,  then the $M$-dependence must also be retained for
consistency.  The $m$-dependence results in the well-known ``slow rescaling"
adjustment of the  scaling variable and  the cross section.
The target mass also adjusts the effective scaling variable, and 
can shift the cross section by up to $25\%$ for fixed-target experiments. 

For collider experiments such as the HERA $e-p$ facility, 
we would like to study charged and neutral current production of 
charm and bottom quarks. Such processes fall in the intermediate region
where the heavy quarks are neither $Q^2\sim m^2$ or $Q^2\gg m^2$;
hence, we must carefully take the mass dependence into account.

In the second paper of this series we shall make use of the framework
developed here  to present a full next-to-leading order analysis of
both charged and neutral current cross sections for deeply inelastic
scattering.


\appendix
\section{
                            Appendix I: Kinematics
\label{app:Kin}
}

We summarize the details about the kinematics including target and
heavy quark mass effects in this appendix.
 We begin with the lab frame kinematics for the overall process, and
then examine the class of colinear frames including the Brick Wall
(BW)  frame. 
 Finally, we consider the colinear frame for the partons, and 
relate the partonic quantities (including dot products) to  the
hadronic variables\rlap.\footnote{ 
We use the metric $g=\{+---\}$ when necessary,  but attempt to
present the results in  a metric independent fashion.}

\subsection{
                                            Overall Process
\label{app:AllKin}
}

For the physical process
\begin{eqnarray}
\ell_1 (\ell_1) + N(P) &\rightarrow &
\ell_2 (\ell_2) + X(P_X)
\end{eqnarray}
the following invariant variables are standard:
\begin{eqnarray}
P^2  &=& M^2
\nonumber \\ [5pt]
Q^2  &=& -q^2
\nonumber \\ [5pt]
\nu &=&
{ P\cdot q \over \sqrt{P\cdot P } } = E_1-E_2
\nonumber \\ [5pt]
x &=&
{ -q^2 \over 2 P\cdot q  } = {Q^2 \over 2 M \nu}
\nonumber \\ [5pt]
y &=&
{ P\cdot q \over P\cdot \ell_1 } =  {\nu \over E_1}
                                                           \label{eq:varDef}
\end{eqnarray}
where $q=\ell_1 - \ell_2$, and
  $E_1$ and $E_2$ are the laboratory energies of the incoming and
outgoing leptons respectively. 

\xfigbasicproc

The components of the relevant
4-vectors in the lab frame are:
\begin{eqnarray}
\begin{array}{lclccr}
P^\mu &= &\Bigl(  M,  &0, &0,  &0 \Bigr)  \\
\ell_1^\mu &= &\Bigl(  E_1,  &0, &0,  &-E_1 \Bigr)  \\
\ell_2^\mu &= &\Bigl(  E_2,  & -E_2 \sin\theta , &0,  &-E_2 \cos\theta \Bigr) \\
q^\mu &= &\Bigl( \nu,  &+E_2 \sin\theta , &0,  &-E_1+E_2 \cos\theta\Bigr)
\end{array} 
               \label{eq:labMom}
\end{eqnarray}
where, as throughout this paper, lepton masses are neglected.

The cross section for the deep inelastic scattering process is given by the
standard form:
\begin{eqnarray}
d\sigma &=&
{1 \over 2\Delta(s,m_{\ell_1}^2,M^2) } \ 
\overline{ \sum_{spin} } \ 
\left| {M}^2  \right| \ 
d\Gamma
                                                            \label{eq:XsecRaw}
\end{eqnarray}
with $M$ being the mass of the incident hadron, $m_{\ell_1}$ the mass of the
incident lepton, and the triangular function
\begin{eqnarray}
\Delta(a,b,c) &=& 
\sqrt{ a^2 + b^2 + c^2 - 2(ab + bc + ca)}
 \label{eq:triangle}
\end{eqnarray}
The sum and average over spins is given by
\begin{eqnarray}
\overline{\sum_{spin}} &=& 
{1\over n_\ell } \ \sum_{spin}
\quad {\rm with} \  n_\ell = 
{\rm \# \ of \ initial \ spin \ states} \ = 
\left\{  
{ 1\ {\rm for}\ \nu,\bar{\nu}  \atop 
  2\ {\rm for}\ \ell^\pm} 
\right.
\ . 
\end{eqnarray}
$d\Gamma$ represents the final state phase space, with all unobserved degrees of
freedom to be integrated over,
\begin{eqnarray}
d\Gamma &=& 
\widetilde{d\ell_2} \ 
(2\pi)^4 \ \delta^4(P+\ell_1 - P_X - \ell_2)\  d\Gamma_X
\end{eqnarray}
with the notation (for invariant single-particle phase space)
\begin{eqnarray}
\widetilde{dk}  &=&
{d^4 k \over (2 \pi)^4 } \ 
(2 \pi) \delta_+(k^2-m_k^2) \  = 
{d^{3} k \over (2 \pi)^{3} (2k_0) } 
\end{eqnarray}
and $d\Gamma_X$ representing the phase space factor for the hadronic final
state. With the scattering amplitude given by \Eq{InvAmplT}, one can put the
various pieces together to get:
\begin{eqnarray}
d\sigma &=&
{G_1 \, G_2  \over 2\Delta(s,m_{\ell_1}^2,M^2) } \ 
4 \pi Q^2 \ 
{L}^{\mu}{}_{\nu} \ 
{W}^{\nu}{}_{\mu} \ 
\widetilde{d\ell_2} \  d\Gamma' 
\end{eqnarray}
Where $G_i = g_B^2/(Q^2+M_{B_i}^2)$, 
the subscripts on $g_{B_i}^2$ and $M_{B_i}^2$ indicate the type of
exchanged vector boson, $d\Gamma '$ represents unintegrated hadron degrees of
freedom (such as those associated with the production of a heavy quark), and the
lepton (hadron) tensor ${L}^{\mu}{}_{\nu}$(${W}^{\nu}{}_{\mu}$) is defined in
\Eq{leptonTen-Def} (\Eq{hadronTen-Def}).
For convenience, ${W}$ and ${L}$ are defined to be dimensionless; these depart
from some historical definitions by simple factors such as $M$.
The factor of $4 \pi Q^2$ comes from the normalization of ${W}$ and ${L}$.

Suppressing $d\Gamma '$, one obtains:
\begin{eqnarray}
{d\sigma \over dx \, dy } &=&
{y Q^2  \over 8 \pi} \ 
G_1 \, G_2   \ 
{L}\cdot {W}
                                                              \label{eq:dsigLW}
\end{eqnarray}
Note that the gauge couplings of the bosons $g_{B_i}$ appear explicitly whereas
the chiral couplings of the leptons $\{ g_{R\, \ell}, g_{L\, \ell} \}$  and
hadrons $\{ g_{R\, h}, g_{L\, h} \}$  are kept with the currents hence reside
in the respective tensors.

For completeness, we record the relations between various commonly used
cross-sections:
\begin{eqnarray}
{d\sigma \over dx \, dy } &=&
2M E_1 x {d\sigma \over dx \, dQ^2 } =
2M E_1^2 y {d\sigma \over dQ^2 \, d\nu } =
{M E_1 y \over E_2} {d\sigma \over dE_2 \, d\cos\theta }
                                                            \label{eq:dsigRel}
\end{eqnarray}
which can be easily derived using the kinematic definitions in \Eq{varDef}.

\subsection{
                                The Colinear Frames
\label{app:ColFrame}
}

Since the underlying physical process is actually the scattering of a
space-like vector-boson on a nucleon, ({\it cf.}, \Fig{basicproc}):
\begin{eqnarray}
B(q) + N(P) &\rightarrow & X(P_X)
\end{eqnarray}
it is more natural to use frames in which the 4-vectors $(q,P)$ define the $t-z$
plane.  For parton-model considerations, it is convenient to specify these
vectors in a general frame of this class by their light-cone coordinate
components $(x^+,\vec{x},x^-)$, with $x^\pm = (x^0\pm x^3)/\sqrt{2}$, as:
\begin{eqnarray}
\begin{array}{lcrll}
P^\mu &= &\Bigl(  P^+,  & \vec0,  & {M^2\over 2 P^+} \Bigr) \\
q^\mu &= &\Bigl(  -\eta P^+,  & \vec0,  & {Q^2\over 2 \eta P^+} \Bigr)
\end{array}
                                                            \label{eq:CLpq}
\end{eqnarray}
where $P^+$ is arbitrary, and $\eta$ is defined through the implicit
equation:
\begin{eqnarray}
2 \, q\cdot P &=& {Q^2 \over \eta} - \eta M^2
\end{eqnarray}
$\eta$ represents the generalization of the familiar Bjorken-$x$ in the presence
of target mass, and it is related to the latter by:
\begin{eqnarray}
{1\over x } &=&
{1\over \eta} - \eta {M^2\over Q^2}
   \label{eq:etax}
\end{eqnarray}
Clearly, $\eta$ reduces to $x$ in the zero target mass limit,
\begin{eqnarray}
\eta &{\longrightarrow \atop M^2/ Q^2 \rightarrow 0 } & x 
\end{eqnarray}
whereas, the general solution to \Eq{etax} is:
\begin{eqnarray}
{1\over \eta } &=&
{1\over2  x } +
\sqrt{ {1\over 4 x^2} + {M^2\over Q^2}   }
   \label{eq:etagen}
\end{eqnarray}
We shall refer to this class of frames as the {\em colinear frames}.  The
laboratory frame (with the negative $z$-axis aligned along $\vec{q}$) belongs to
this class; it is obtained by setting $P^+=M/\sqrt{2}$.  The ``infinite momentum
frame," often used to derive the QCD asymptotic theorems, is obtained in the
limit $P^+\rightarrow \infty$.  Another useful frame in this class, used in the
helicity formulation, is discussed in the following.

\subsection{
                                         The Brick Wall Frame
\label{app:BWframe}
}

The Brick Wall (BW) frame is the natural ``rest-frame" of the exchanged vector
boson when its momentum $q$ is space-like, $q^2 = -Q^2 <0$,
({\it cf.} \Fig{lhProc}). It is also one of the colinear frames---corresponding
to setting $P^+=Q/(\eta \sqrt{2})$ in \Eq{CLpq}, hence obtaining $q^0=0$ and
$q^3=-Q$.  
 In the cartesian coordinate system, $(x^0, x^1, x^2, x^3 )$,  we have: 
\begin{eqnarray}
\begin{array}{lcclccr}
q^\mu   &= &Q           &\Bigl(  0,  &0, &0,  & -1 \Bigr)  \\
P^\mu   &= &{1\over 2Q} &\Bigl(  \Delta_P,  &0, &0,  & +\beta_1\Bigr)  \\
P_X^\mu &= &{1\over 2Q} &\Bigl(  \Delta_P,  &0, &0,  & -\beta_2\Bigr)  \\
\end{array}
         \label{eq:BWqpx}
\end{eqnarray}
and we refer to this frame as the 
{\it standard hadron configuration}, \Fig{BW2},
with
\begin{eqnarray}
\Delta_P &=&  \Delta[-Q^2,P^2,P_X^2]  \nonumber \\
\beta_1 &=&  Q^2-P^2+P_X^2  \nonumber \\
\beta_2 &=&  Q^2+P^2-P_X^2
\end{eqnarray}

\xfigBWii

In this frame,the lepton momenta are given by:
\begin{eqnarray}
\begin{array}{lcclccr}
\ell_1^\mu &= &{Q\over 2} &\Bigl(\cosh\psi,  &\sinh\psi, &0,  &-1 \Bigr)  \\
\ell_2^\mu &= &{Q\over 2} &\Bigl(\cosh\psi,  &\sinh\psi, &0,  &+1 \Bigr)  \\
\end{array}
      \label{eq:BWl12}
\end{eqnarray}
which can be easily envisioned as being obtained from the {\em standard lepton
configuration} ({\it cf.} the standard hadron configuration, \Eq{BWqpx}),
\Fig{BW1},
\begin{eqnarray}
\begin{array}{lcclccr}
\ell_1^\mu &= &{Q\over 2} &\Bigl(1,  &0, &0,  &-1 \Bigr)  \\
\ell_2^\mu &= &{Q\over 2} &\Bigl(1,  &0, &0,  &+1 \Bigr)  \\
\end{array}
\label{eq:StdBW}
\end{eqnarray}
by a ``rotation'' in the $(t-x)$ plane (really a Lorentz boost)  by the
hyperbolic angle $\psi$.  This is in analogy to the familiar CM rotation [in the
$(z-x)$ plane] between initial and final scattering states in a time-like
situation.  This is illustrated in \Fig{BW1} and \Fig{BW2}.

\xfigBWi

The hyperbolic cosine can be obtained from the formula:
\begin{equation}
\cosh\psi =  \frac{2 P \cdot (\ell_1+\ell_2)} {\Delta[-Q^2,P^2,P_X^2]}  
\end{equation}
Evaluating the scalar productions in the laboratory frame, we relate $\cosh\psi$ to
the more familiar variables:
\begin{eqnarray}
\cosh\psi &=& 
{ E_1+E_2   \over \sqrt{ Q^2+\nu^2}} 
=
{\eta^2 M^2 - Q^2 + 2 \eta (s-M^2) \over \eta^2 M^2 + Q^2 } 
\qquad 
{ \longrightarrow \atop M\rightarrow 0 } 
\qquad 
{(2-y) \over y}
                   \label{eq:coshPsi}
\end{eqnarray}

In developing the helicity formalism (\App{SF}), we encounter the ``spin-1
rotation matrix'' for the vector boson polarization vectors under the above
Lorentz boost from the configuration \Eq{StdBW} (\Fig{BW1}) to \Eq{BWl12}
(\Fig{BW2}).  The 3-dimensional d-matrix is: 
\begin{eqnarray}
d^1(\psi) &=& 
\left[ \begin{array}{ccc}
{1+\ch\psi\over2}& {-\sh\psi\over\sqrt{2}} & {1-\ch\psi\over2}\\ [10pt] 
{-\sh\psi\over\sqrt{2}} & \ch\psi& {+\sh\psi\over\sqrt{2}}  \\  [10pt] 
{1-\ch\psi\over2}& {+\sh\psi\over\sqrt{2}} & {1+\ch\psi\over2}
\end{array}\right]  
                                                            \label{eq:d-matrix}
\end{eqnarray}
It is the SO(2,1) analogue of the familiar SO(3) rotation matrix.

\subsection{
                        Parton Kinematics in the QCD Parton Model
\label{app:ParKin}
}

In the QCD Parton Model ({\it cf.}, \Fig{FactThm}), we have an initital state
parton momentum $k_1$, whose light-cone components in a colinear frame are:
\begin{eqnarray}
\begin{array}{lcrll}
k_1^\mu &= &\Bigl(  \xi P^+,  & \vec0,  & {m_1^2\over 2\xi P^+} \Bigr) \\
\end{array}
\end{eqnarray}
where $\xi$ is the fractional momentum carried by the parton. The momenta involved
in the ``hard scattering'' consist of
\begin{equation}
q + k_1 \rightarrow k_x
\end{equation}
where the final state, represented by the total momentum $k_x$, consists of either
a on-mass-shell single parton (for the case of the LO calculation) or a continuum of
multi-parton configurations (for the NLO calculations and beyond).  
 
For the LO calculation presented in \Sec{LO}, with $k_x=k_2=k_1+q$, we
can evaluate the argument of the delta function which enforces the on-shell
condition for the final state heavy quark:
\begin{equation}
k_2^2 -m_2^2  = \frac{Q^2 (\xi - \chi_+)(\xi - \chi_-)}{\eta \xi} 
\label{eq:hqkMsShll}
\end{equation}
where 
\begin{eqnarray}
\chi_\pm  &=&  \eta \ 
{ (Q^2-m_1^2+m_2^2) \pm \Delta[-Q^2,m_1^2,m_2^2]  \over 2 Q^2 }
\end{eqnarray}
and $\eta$ is defined in \Eq{etagen}.
The limits on $\xi$ (see below) dictate that the only physical root is
\begin{equation} 
\xi = \chi \equiv \chi_+ 
                                                            \label{eq:xiChi}
\end{equation}
This variable reduces to the  ``slow-rescaling'' variable 
$x(1+m_2^2/Q^2)$ in the limit  $m_1\to 0$ {\it and} $M\to 0$.
Substituting \Eq{xiChi} in the second factor in \Eq{hqkMsShll}, we obtain
\begin{eqnarray}
\delta_+(k_2^2 -m_2^2)  &=&
{ \delta\left(  {\xi\over\chi} - 1 \right)
\over  \Delta[-Q^2,m_1^2,m_2^2] }
\end{eqnarray}

	   When the final state consists of multi-partons (for NLO and beyond), the CM
energy of the subprocess $\hat{s}$ must be greater than a threshold
$\hat{s}_{th}$, which is  equal to either $m_2^2$ or $4m_2^2$, depending on whether
a single heavy quark (charged current case) or a heavy quarks-antiquark pair(neutral
current case) is produced.  Since 
\begin{eqnarray}
\hat{s} &=& (k_1+q)^2 = m_1^2 -Q^2 + 2k_1\cdot q =
\left( Q^2 + \frac{\eta}{\xi} \, m_1^2 \right) 
\left(  \frac{\xi}{\eta} - 1 \right) 
\geq \hat{s}_{th}
\end{eqnarray}
it is easy to see that the threshold condition imposes
the constraint $\xi \geq \xi_{th}$ on the parton momentum fraction variable where
\begin{eqnarray}
\xi_{th}   &=&  \eta \ 
{ (Q^2-m_1^2+\hat{s}_{th}) + \Delta[-Q^2,m_1^2,\hat{s}_{th}]  \over 2 Q^2 }
\end{eqnarray}
(Note that for $\hat{s}_{th} = m_2^2$,  $\xi_{th} = \chi_+ \equiv \chi$.) 
On the other hand, the condition that $P_X^+= P^+(1-\xi) \geq 0$ requires $\xi
\leq 1$.  Hence, $\xi$, which is also the integration variable for the convolution
in the fundamental factorization theorm (\Eq{FactThm:Wmunu}), has the following range:
\begin{eqnarray}
1 \geq \xi \geq \xi_{th} &=& 
 \eta \ 
{ (Q^2-m_1^2+\hat{s}_{th}) + \Delta[-Q^2,m_1^2,\hat{s}_{th}]  \over 2 Q^2 }
            \label{eq:xiRange}
\end{eqnarray}
We recall that $\eta$ is the generalization of Bjorken $x$ incorporating the target
mass effect.  Thus the lower limit for $\xi$ is modified by both target mass {\it and}
heavy quark mass.  This aspect of mass-dependence has been overlooked in existing
literature.

\subsection{
                                  Dot Productions of Lepton and Parton Momenta
\label{app:lepDotHad}
}

In the explicit calculation of cross-sections using the contraction of lepton
and hadron tensors, (\cf, \Sec{LO} and \App{LO}) one needs the scalar products of
the lepton and hadron 4-vectors.  
 This calculation is subtle because the variable $\xi = k_1^+ / P_+$ is 
invariant for boosts along the $z$-axis, but not for 
other boosts or rotations. 

In the BW frame, the light-cone components of the two parton momenta are:
\begin{eqnarray}
\begin{array}{lccrlclr}
k_1^\mu  &: & \frac{Q}{\sqrt{2}}
            &\Bigl( 
            & \frac{\xi}{\et} ,
	    &\vec{0}, 
            &\frac{\et}{\xi}  \frac{m_1^2}{Q^2}  
            & 	  \Bigr)  
\\ [10pt]
k_2^\mu  &: & \frac{Q}{\sqrt{2}} 
            &\Bigl( 
       	    &\frac{\xi-\eta}{\eta}   ,
	    &\vec{0}, 
            & 1 + \frac{\eta}{\xi} \frac{m_1^2}{Q^2} ,
            &   \Bigr)  
\\
\end{array}
                 \label{eq:BWk12}
\end{eqnarray}

Using the explicit components of the lepton momenta given in \Eq{BWl12},
it is then straightforward to show
\begin{eqnarray}
(k_1 \cdot \ell_1) &=&
{1\over2}\ {\xi \over \eta}\ Q^2   \left( {\cosh\psi+1 \over 2}  \right) +
{1\over2}\ {\eta \over \xi}\ m_1^2 \left( {\cosh\psi-1 \over 2}  \right)
\label{eq:k1l1i}
\end{eqnarray}
\begin{eqnarray}
(k_1 \cdot \ell_2) &=&
{1\over2}\ {\xi \over \eta}\ Q^2   \left( {\cosh\psi-1 \over 2}  \right) +
{1\over2}\ {\eta \over \xi}\ m_1^2 \left( {\cosh\psi+1 \over 2}  \right)
\label{eq:k1l2i}
\end{eqnarray}
To contrast the simplicity and symmetry of this group theoretic approach 
with a more traditional ``brute force" calculation {\it in the colinear frame}, 
we compare: 
\begin{eqnarray}
(k_1\cdot\ell_1) &=&
{ Q^2 \xi^2 ( s - M^2 + M^2 \eta) + m_1^2 (s \eta^2 - M^2 \eta^2 -Q^2\eta)
\over  2\xi (Q^2 + M^2 \eta^2) }
\label{eq:k1l1ii}
\end{eqnarray}
\begin{eqnarray}
(k_1\cdot\ell_2) &=&
{ Q^2 \xi^2 ( s - M^2 -Q^2/\eta) + m_1^2 \eta^2 (s  + M^2 \eta -M^2)
\over  2\xi (Q^2 + M^2 \eta^2) }
\label{eq:k1l2ii}
\end{eqnarray}
Although it is not obvious, \Eq{k1l2i} and \Eq{k1l2i}  are identical to \Eq{k1l2ii}
and \Eq{k1l2ii}; however, the symmetries of the problem are more apparent in 
\Eq{k1l2i} and \Eq{k1l2i}.

In the limit of zero masses, we have the usual relations where
$(k_1\cdot\ell_1) \rightarrow \hat{s}/2 $ and
$(k_1\cdot\ell_2) \rightarrow \hat{u}/2 $ with no $\xi$ dependence.
However, if we wish to obtain the correct mass dependence, we must
include the proper $\xi$ dependence in our calculation.           

Once we have $(k_1\cdot\ell_1)$ and $(k_1\cdot\ell_2)$, we can use
$k_1+\ell_1=k_2+\ell_2$ to easily compute the other necessary
combinations via:
\begin{eqnarray}
(k_2 \cdot \ell_2) &=& (k_1 \cdot \ell_1) -
\left( {m_2^2-m_1^2 \over 2} \right)
\nonumber\\ [10pt]
(k_2 \cdot \ell_1) &=& (k_1 \cdot \ell_2) +
\left( {m_2^2-m_1^2 \over 2} \right)
\end{eqnarray}
%


\section{
                    Appendix II: Structure Functions and Cross-sections
\label{app:SF}
}

Since the precise treatment of the mass effects is emphasized in this paper, we
include here some details on the derivation of structure function and
cross-section formulas used in the text, especially for the less familiar
helicity vertices and structure functions.

\subsection{
             Tensor Amplitudes and Invariant Structure Functions
\label{app:TenAmp}
}

We begin by recording the expression for the lepton tensor, \Eq{leptonTen-Def}.
In the limit of zero lepton mass, it is:

\begin{eqnarray}
{L}^{\mu\nu}  &=&
{1\over Q^2} \ 
\overline{\sum_{spin}} \ 
\overline{u}(\ell_1) \, \Gamma^\mu          \, u(\ell_2) \cdot
\overline{u}(\ell_2) \, \Gamma^{\nu\dagger} \, u(\ell_1) 
=
{1\over Q^2} \ {1 \over n_\ell} \ 
{\rm Tr}[\ell_1  \, \Gamma^\mu  \, \ell_2  \Gamma^{\nu\dagger} ]
\end{eqnarray}
where $n_\ell$ counts the number of incoming helicity states. Using a general
V-A coupling of the form, \Eq{lepCurT},
\begin{eqnarray}
 \Gamma^\mu &=&
\gamma^\mu \, [g_{R\, \ell} (1+\gamma_5) + g_{L\, \ell} (1-\gamma_5) ] 
\end{eqnarray}
the result is:
\begin{eqnarray}
{L}^{\mu\nu}  &=&
{8\over Q^2} \ {1\over n_\ell}  \ 
\Bigl\{
g_{+ \, \ell}^2 
      \left[ \ell_1^\mu  \ell_2^\nu + \ell_2^\mu  \ell_1^\nu 
       - g^{\mu\nu} \, {Q^2\over2} \right] 
-g_{- \, \ell}^2 
    \left[ i \epsilon^{\mu\nu\rho\sigma} \ell_{1\rho}  \ell_{2\sigma} 
 \right]
\Bigr\}  \nonumber\\
                                                              \label{eq:apLepTen}
\end{eqnarray}

The independent components of the hadron tensor $W_{\mu\nu}$ are expressed in
terms of invariant ({\it i.e.}, Lorentz scalar) structure functions defined as
(\Eq{invStrFun}),
\begin{eqnarray}
    W^{\mu}{}_{\nu}
        & = &
      - g^{\mu}{}_{\nu} W_{1}
      + \frac{P^{\mu} P_{\nu}}{M^2}  W_{2}
      - i \frac{\epsilon^{Pq\mu}{}_{\nu}}{2 M^2} W_{3} + \ \ \ \label{eq:invSF}
\\ & & \makebox[.10in]{}
      + \frac{q^{\mu} q_{\nu}}{M^2} W_{4}
      + \frac{P^{\mu} q_{\nu} + q^{\mu} P_{\nu}}{2 M^2} W_{5}
      + \frac{P^{\mu} q_{\nu} - q^{\mu} P_{\nu}}{2 M^2} W_{6}
\nonumber                                                      
\end{eqnarray}

Contracting the lepton and hadron tensors and
evaluating the scalar productions of the 4-vectors in the laboratory frame 
(\cf, \Eq{labMom}), one obtains:
\begin{eqnarray}
{W} \cdot  {L}  &=&
{16 E_1 E_2 \over {n_\ell Q^2}} \ 
\left\{
g_{+ \, \ell}^2  \left[ 2 \, \sin^2{\theta\over2} \, W_1 + 
         \cos^2  {\theta\over2} \, W_2\right] 
+ g_{- \, \ell}^2 
    \left[{E_1+E_2\over M} \, \sin^2{\theta\over2} \, W_3 \right] 
\right\}
\nonumber\\                                               \label{eq:LdotW}
\end{eqnarray}
The structure functions $\{W_4,W_5,W_6\}$ do not appear on the right-hand side
of this equation because the dot product of $q^\mu$ with the lepton tensor
${L}^{\mu\nu}$ gives rise to a factor proportional to some combinations of the
lepton masses which is neglect here.  \Eq{LdotW}, in conjunction with
\Eqs{dsigLW}{dsigRel}, form the bases for the derivation of the cross-section
formula \Eq{dSigma/dEdz} in \Sec{StrFun}.

\subsection{
                    Helicity Vertices and Structure Functions
\label{app:HelStrFun}
}

We now turn to the calculation of helicity amplitudes, vertices, and structure
functions.  We use the helicity labels $\lambda_{1,2}$ for the leptons;
$\sigma_{1,2}$ for the hadrons, and $\{ m,n \}$ for the bosons.  Lower indices
are for incoming particles; and upper indices are for outgoing particles.  The
scattering amplitude for the basic process, \Eq{lhProc}, can be written in the
factorized form in the helicity basis:\rf{helForm1,helForm2}
\begin{eqnarray}
{\cal M}_{\lambda_1 \, \sigma_1}^{\lambda_2 \, \sigma_2} &=&
J_{\sigma_1 \, m}^{* \, \sigma_2} (Q^2, q\cdot P) \ 
{g_B^2 \ d^1(\psi)^m{}_n \over Q^2 + M_B^2 } \ 
j_{\lambda_1}^{\lambda_2 \, n} (Q^2) \ 
                                                          \label{eq:apAmpH}
\end{eqnarray}
where $d^1(\psi)^m{}_n$ is a spin-1 $SO(2,1)$ ``rotation matrix'' in the
Brick-Wall frame of the process  corresponding to
$q^\mu~:(~0,~0,~0,~-Q$), 
({\it cf.},\Eq{d-matrix}). The scalar lepton helicity vertex function is:
\begin{eqnarray}
j_{\lambda_1}^{\lambda_2 \, n} (Q)  &=&
\epsilon^{n \, *}_{\mu} 
\langle \ell_2 , \lambda_2 | j^\mu  | \ell_1 , \lambda_1 \rangle 
\quad = \quad 
\overline{u}_{\lambda_2}(\ell_2) \ 
\epsilon^{n \, *}\cdot \Gamma \ u_{\lambda_1}(\ell_1) 
\end{eqnarray}
and the corresponding hadron vertex function is:
\begin{eqnarray}
J_{\sigma_1 \, m}^{* \, \sigma_2} (Q^2, q\cdot P)  &=&
\langle P_X ,  \sigma_2 | J_{\mu}^{\dagger} | P ,\sigma_1 \rangle \, 
\epsilon_m^\mu
\end{eqnarray}
Much of the simplicity of the helicity approach results from the fact that the
lepton vertex function is extremely simple in the limit of zero lepton masses.
{\em For left-handed (right-handed) coupling, there is only one non-vanishing
vertex function for which all three particles are left-handed (right-handed)};
it is simply given by:
\begin{equation}
j_{L}^{L \, L} (Q) = j_{-1/2}^{-1/2 \, -1} (Q) = \sqrt{8Q^2}
\end{equation}
(Likewise, $j_{R}^{R \, R} (Q) = -\sqrt{8Q^2}$ in the case of right-handed
coupling).  Thus, upon squaring the scattering amplitude, \Eq{apAmpH}, one
obtains:
\begin{equation}
\overline{ \sum_{spin} } \ \left| {M}^2  \right| \ \propto \ 
d^1(\psi)^{-1}{}_{m} \ \  d^1(-\psi)^{n}{}_{-1} \ \ {W}^{m}{}_{n} \ 
\end{equation}
where ${W}^{m}{}_{n}$ is the helicity forward Compton scattering amplitude for
initial state vector boson polarization $n$ and final state polarization $m$:
\begin{equation}
    {W}^{m}{}_{n}
        =
        \epsilon^{m}_{\mu}{}^{*} (P,q) \, 
        W^{\mu}{}_{\nu} (P,q) \, 
        \epsilon_{n}^{\nu} (P,q)
                                                      \nonumber
\end{equation}
For totally inclusive process, this amplitude must be diagonal in $(m,n)$ due to
angular momentum conservation;\footnote{In principle, there can be mixing
amoung  $\{ W^{q}{}_{q}, W^{q}{}_{0}, W^{0}{}_{q} \}$. Since the
coefficients of these terms are proportional to $m_{\ell}^2/Q^2$,
we only concern ourselves with $\{ W^{+}{}_{+}, W^{0}{}_{0}, W^{-}{}_{-} \}$.}
hence, the right-hand side becomes
$d^1(\psi)^{-1}{}_{m} \ d^1(\psi)^{m}{}_{-1} \ F_{m}$ where the diagonal
helicity amplitude ${W}^{m}{}_{m}$ is identified with the {\em helicity
structure function $F_{m}$}, \cf, \Eq{StrFun-Hel}.

Using these results for the squared amplitude, $|M^2|$, keeping all factors,
and making use of the explicit form of the d-matrix, \Eq{d-matrix}, we obtain
${L} \cdot {W}$, which appears in the cross-section formula \Eq{dsigLW}: 
\begin{eqnarray}
{W} \cdot {L}  = {8\over n_\ell}                          
\Bigg\{ &g_{R\, \ell}^2 &
\left[
 F_+ \left( {1+\cosh\psi \over 2} \right)^2 
+F_0 \left( {-\sinh\psi \over \sqrt{2}} \right)^2 
+F_- \left( {1-\cosh\psi \over 2} \right)^2 
\right] 
\nonumber                                                              
\\+ &g_{L\, \ell}^2 &
\left[
 F_+ \left( {1-\cosh\psi \over 2} \right)^2 
+F_0 \left( {+\sinh\psi \over \sqrt{2}} \right)^2 
+F_- \left( {1+\cosh\psi \over 2} \right)^2 
\right] 
\Bigg\}
\label{eq:WdotL-H}
\end{eqnarray}
This leads to the general formula, \Eq{dSigma-Hel}, for the cross-section
given in \Sec{StrFun}.

\subsection{
             Relations between Invariant and Helicity Structure Functions
\label{app:Inv-Hel}
}

To derive the relations between the invariant and helicity structure
functions, we first examine the polarization vectors for a vector boson with
momentum $q$ in the helicity basis.  With respect to an arbitrary reference
momentum $p$, the ``longitudinal'' polarization vector is:
\begin{equation}
\epsilon_0^\mu (p,q) =
{ (-q^2) \, p^\mu  + (p\cdot q) \, q^\mu \over 
     \sqrt{(-q^2)[(p\cdot q)^2 - q^2 p^2]} }
           \label{eq:longVec}
\end{equation}
with $-q^2=Q^2>0$ for space-like $q^\mu$. 
It is also useful to define the ``scalar" polarization: 
\begin{equation}
\epsilon_q^\mu (p,q) =
{ q^\mu  \over \sqrt{-q^2} }
    \label{eq:qVec}
\end{equation}
In a collinear frame where the z-component of $q^\mu$ is positive, the
transverse polarization vectors are given by:
\begin{eqnarray}
\epsilon_\pm^\mu ( p,q) &=&   { 1\over \sqrt{2} } ( 0, \mp 1, -i, 0) 
                                                        \label{eq:tranVec}
\end{eqnarray}
 For the z-component of $q^\mu$ negative, we rotate the above about the $y$-axis
by $\pi$. 
 These polarization vectors depend on the reference vector $p^\mu$
only to the extent that it defines the $t-z$ plane in conjunction with
$q^\mu$.  For the transverse polarization vectors, this is
obvious.  For the longitudinal vector, $\epsilon_0^\mu ( p,q)$, this follows from
the fact that it is merely the unit vector in the $t-z$ plane orthogonal to
$q^\mu$. The reference vector $p^\mu$ is used only to define this plane and to
provide the non-vanishing perpendicular component for projecting onto
$\epsilon_0^\mu$. The two distinct reference vectors in the plane, such as
$P^\mu$ (the target momentum) and $k_1^\mu$ (the initial state parton momentum)
used in the text, define the same set of polarization vectors for the vector
boson.  As discussed in \Sec{FacThm}, this is the key point which leads to the
simple factorization formula for the helicity structure functions in the QCD
Parton framework.

To project out the transverse helicity amplitudes, the following representations 
are useful:
\begin{eqnarray}
\epsilon_+^{\mu} (p,q) \  \epsilon_+^{\nu *} (p,q) \  - 
\epsilon_-^{\mu} (p,q) \  \epsilon_-^{\nu *} (p,q) \   
&=&
{ i \epsilon^{\mu \nu p q} \over \sqrt{ (p\cdot q)^2 - q^2 p^2  }  }
\nonumber \\
\epsilon_+^{\mu} (p,q) \  \epsilon_+^{\nu *} (p,q) \  + 
\epsilon_-^{\mu} (p,q) \  \epsilon_-^{\nu *} (p,q) \   
&=&
- g^{\mu \nu} 
+ \epsilon_0^{\mu} (p,q) \  \epsilon_0^{\nu *} (p,q) \
- \epsilon_q^{\mu} (p,q) \  \epsilon_q^{\nu *} (p,q) 
    \label{eq:pmVec}
\end{eqnarray}
The second relation is simply completeness. 

Applying the above polarization vectors to the definition of the helicity
structure functions, \Eq{StrFun-Hel},
\begin{equation}
    F_{\lambda}
        =
        \epsilon^{\lambda}_{\mu}{}^{*} (P,q) \, 
        W^{\mu}{}_{\nu} (P,q) \, 
        \epsilon_{\lambda}^{\nu} (P,q)
   \makebox[.5in]{} ({\rm no\ sum\ over\ }\lambda)
                                                      \nonumber
\end{equation}
and using the representation of $W^{\mu}{}_{\nu} (P,q)$ in terms of the
invariant structure functions, \Eq{invSF}, we obtain:
\begin{eqnarray}
\begin{array}{rcrcl}
F_+  &=&  W_1  &-& {\nu \over 2M  }\ \sqrt{1 + {Q^2 \over \nu^2 } }\  W_3   
  \\ [5pt]
F_-  &=&  W_1  &+& {\nu \over 2M  }\ \sqrt{1 + {Q^2 \over \nu^2 } }\  W_3   
  \\ [5pt]
F_0  &=& -W_1  &+& \left( 1 + {\nu^2 \over Q^2 } \right)  \ W_2   
\end{array} 
   \label{apFvsW}
\end{eqnarray}
The complete transformation matrix to convert hadron helicity amplitudes to 
invariant  amplitudes
($W_\lambda = f \otimes \omega_{\lambda} = {t\,}_{\lambda}^{i} \,  W_i$)
is given in \Tbl{wlamtowi}. 
The coefficients for the inverse transformation, $(t^{-1})^{\, \lambda}_{\, i}$,
are given in \Tbl{witowlam}.

\subsection{Relations Between Hadron and Parton Tensors}

As discussed in \Sec{FacThm}, the  $k_1$ 4-vector is not simply proportional to
$P$, but in general contains a mixture of $P$ and $q$ given by:
\begin{eqnarray}
    k_1^{\mu} &=& \zeta_{P} P^{\mu} + \zeta_{q} q^{\mu}
\nonumber\\[5pt]
\zeta_{P} &=& { Q^2 \xi^2 +m_1^2 \eta^2 \over \xi (Q^2 +M^2 \eta^2) }
\\[5pt]
\zeta_{q} &=& { \eta( m_1^2 -M^2 \xi^2 ) \over \xi (Q^2 +M^2 \eta^2) }
\nonumber
\label{eq:kqPrel} 
\end{eqnarray}
Note that this mixing depends on both $M$ and $m_1$. 
The result is that the hadron tensors and the parton tensors are mixed. 
Specifically, 
\begin{eqnarray}
W_i &=& 
c_{\, i}^{\, j} \  f \otimes \omega_{j}
\end{eqnarray}
where the $c_{\, i}^{\, j}$ coefficients are given in \Tbl{omgtow}.
The coefficients for the inverse transformation, $(c^{-1})^{\, i}_{\, j}$,
are given in \Tbl{wtoomg}.

This is in contrast to the corresponding result for the hadron helicity
amplitudes where there is no mixing: 
\begin{eqnarray}
F_\lambda &=&  W_{\lambda \, \lambda} = f \otimes \omega_{\lambda}
\end{eqnarray}
%


%

\newpage
\clearpage

\begin{table}[h]
\bigskip
\[
\begin{array}{c|cccccc}
t_{\, \lambda}^{\, i} 
  & F_1 \equiv W_1 & F_2\equiv(\nu/M)W_2  & F_3 \equiv (\nu/M)W_3
& W_4 & W_5 & W_6  \\ \hline \txstrut
F_{+} \equiv W_{++} & 1 & 0 & \frac{-\rho}{2} & 0 & 0 & 0 
\\ \txstrut
F_{-} \equiv W_{--}& 1 & 0 & \frac{+\rho}{2} & 0 & 0 & 0 
\\ \txstrut
F_{0} \equiv W_{00}&-1 &  \frac{\rho^2}{2x}   & 0 & 0 & 0 & 0 
\\ \txstrut
W_{qq} & 1 & \frac{1}{2x} & 0 & \frac{2 Q^2}{M^2} & \frac{-\nu}{M} & 0 
\\ \txstrut
W_{0q}+W_{q0}  & 0 & \frac{\rho}{x} & 0 & 0 & \frac{-\rho \nu}{M} & 0                              
\\ \txstrut
W_{0q}-W_{q0}  & 0 & 0 & 0 & 0 & 0 & \frac{-\rho \nu}{M} \\ 
\end{array}
\]
\caption{ 
  Transformation matrix to convert hadron helicity amplitudes to invariant
  amplitudes: 
  $W_{\lambda \, \lambda} = f \otimes \omega_{\lambda}  =
  {t\,}_{\lambda}^{i} \,  W_i$. 
  Note, we use the short hand notation 
  $F_{\lambda} \equiv W_{\lambda\lambda}$.  
  We have defined $\rho^2=1+Q^2/\nu^2$, 
  and we have $\rho\to 1$ in the DIS limit.
\label{tab:wlamtowi}  
}
\end{table}

\begin{table}[h]
\bigskip
\[
\begin{array}{c|cccccc}
(t^{-1})^{\, \lambda}_{\, i} & F_{+} \equiv W_{++} & F_{-} \equiv W_{--} &
F_{0} \equiv W_{00}  & W_{qq} & (W_{0q}+W_{q0}) & (W_{0q}-W_{q0})  
\\ \hline \txstrut 
F_1 \equiv W_1 & \frac{1}{2} & \frac{1}{2} & 0 & 0 & 0 & 0 \\ \txstrut
F_2\equiv(\nu/M)W_2 
      & \frac{x}{\rho^2} & \frac{x}{\rho^2} & 
        \frac{2x}{\rho^2}  & 0 & 0 & 0 \\ \txstrut
F_3 \equiv (\nu/M)W_3
      & \frac{-1}{\rho}  & \frac{+1}{\rho} & 
            0 & 0 & 0 & 0 \\ \txstrut 
W_{4} & \frac{-M^2}{ 4 \nu^2 \rho^2} & \frac{-M^2}{  4 \nu^2 \rho^2}  & 
        \frac{M^2}{ 2 Q^2 \rho^2}  & \frac{M^2}{ 2 Q^2}  & 
        \frac{-M^2}{ 2 Q^2 \rho}  & 0 \\ \txstrut
W_{5} & \frac{M}{\nu \rho^2} & \frac{M}{\nu \rho^2} & 
        \frac{2M}{\nu \rho^2}  & 0 & \frac{-M}{\nu \rho}  & 
                0 \\ \txstrut       
W_{6} & 0 & 0 & 0 & 0 & 0 & \frac{-M}{\nu \rho} \\ 
\end{array}
\]
  \caption{ 
  Transformation matrix to convert hadron invariant amplitudes to helicity
  amplitudes: $W_i = (t^{-1})^{\, \lambda}_{\, i} \,  W_\lambda$. Note, we use
  the short hand notation $F_{\lambda} \equiv W_{\lambda\lambda}$.
   We have also used $F_1=W_1$,  $F_2=(\nu/M)W_2$, and  $F_3=(\nu/M)W_3$.
  We have defined $\rho^2=1+Q^2/\nu^2$, 
  and we have $\rho\to 1$ in the DIS limit.
  Note that as $M\to 0$, $\{W_4, W_5, W_6\}$ decouple from 
  $\{F_+, F_0, F_- \}$. 
\label{tab:witowlam}  
}
\end{table}

\newpage
\clearpage
\begin{table}[h]
\bigskip
\[
\begin{array}{c|cccccc}
c^{\, j}_{\, i} & \omega_{1} & \omega_{2} & \omega_{3} 
& \omega_{4} & \omega_{5} & \omega_{6}   \\ \hline \txstrut
 F_1 \equiv W_1 & 1 & 0 & 0 & 0 & 0 & 0 \\ \txstrut
F_2\equiv(\nu/M)W_2 & 0 & \frac{\zeta_{P}^2}{2x} & 0 & 0 & 0 & 0 \\ \txstrut
F_3 \equiv (\nu/M)W_3 & 0 & 0 & \frac{\zeta_{P}}{2x}  & 0 & 0 & 0 \\ \txstrut
W_{4} & 0 & \frac{\zeta_{q}^2 M^2}{Q^2}  & 0 & \frac{M^2}{Q^2}  & 
             \frac{\zeta_{q} M^2}{Q^2}  & 0 \\ \txstrut
W_{5} & 0 & \frac{2  \zeta_{P} \zeta_{q} M^2}{Q^2}  & 0 & 0 & 
            \frac{\zeta_{P} M^2}{Q^2}  & 0 \\ \txstrut
W_{6}& 0 & 0 & 0 & 0 & 0 & \frac{\zeta_{P} M^2}{Q^2}  \\ 
\end{array}
\]
  \caption{ 
Transformation matrix to convert parton invariant amplitudes to 
hadron invariant
amplitudes: $W_i= c_{\, i}^{\, j} \  f \otimes \omega_{j}$. 
Note that as $M\to 0$, $\{W_4, W_5, W_6\}$ decouple from 
$\{\omega_i\}$. 
\label{tab:omgtow}  
}
\end{table}

\begin{table}[h]
\bigskip
\[
\begin{array}{c|cccccc}
(c^{-1})_{\, j}^{\, i} 
  & F_1 \equiv W_1 & F_2\equiv(\nu/M)W_2 & F_3 \equiv (\nu/M)W_3 
& W_{4} & W_{5} & W_{6}   \\ \hline \txstrut
\omega_{1} & 1 & 0 & 0 & 0 & 0 & 0 \\ \txstrut
\omega_{2} & 0 & \frac{2x}{\zeta_{P}^2} & 0 & 0 & 0 & 0 \\ \txstrut
\omega_{3} & 0 & 0 & \frac{2x}{\zeta_{P}} & 0 & 0 & 0 \\ \txstrut
\omega_{4} & 0 & \frac{2x \zeta_{q}^2}{\zeta_{P}^2} & 0 & 
       \frac{Q^2}{M^2} & \frac{-\zeta_{q} Q^2}{\zeta_{P} M^2} & 0 \\ \txstrut
\omega_{5} & 0 & \frac{-4x \zeta_{q}}{\zeta_{P}^2} & 0 & 0 & 
       \frac{Q^2}{\zeta_{P} M^2} & 0 \\ \txstrut
\omega_{6} & 0 & 0 & 0 & 0 & 0 & \frac{Q^2}{\zeta_{P} M^2} \\ 
\end{array}
\]
  \caption{ 
Transformation matrix to convert hadron invariant amplitudes to parton
invariant amplitudes: $\omega_j= (c^{-1})^{\, i}_{\, j} \  f \otimes W_{i}$. 
\label{tab:wtoomg}  
}
\end{table}



\section{
                 Leading Order Calculation with Masses
\label{app:LO}
}

We present the details of the leading order calculation with the full mass
dependence both as an illustration of general points made in the text of the
paper, and as a
concrete example to check the self-consistency of the tensor and helicity
formalisms developed in the text.  Although the
calculation is straightforward, the results with the full mass dependence do not
exist in the literature, and have not being used in the analysis of experimental
data---as emphasized in this paper.

The {\em parton structure function} tensor $\omega_{\mu\nu}^{Ba}$, representing
the vector boson ($B$) and parton ($a$) forward Compton scattering amplitude, is
entirely analogous to $W_{\mu \nu}^{BN}$---replacing the hadron target $N$ by
the parton target $a$. The leading order diagram, \Fig{Born}, gives rise to:
\begin{eqnarray}
\omega^\mu_\nu  &=&
{1\over 4 \pi} \ (2\pi)\delta_+(k_2^2-m_2^2) 
\overline{\sum_{spin}} \ 
\langle k_1 , \sigma_1 | j^\mu   | k_2 , \sigma_2 \rangle 
\langle k_2 , \sigma_2 | j^*_\nu | k_1 , \sigma_1 \rangle  \ 
           \label{eq:loOmg}
\end{eqnarray}
For quarks, the spin sum/average on the right-hand-side is:
\begin{eqnarray}
{1\over 2} {\rm Tr} [(\rlap/k_1 + m_1)  \Gamma^\mu 
(\rlap/k_2 + m_2)  \Gamma^{\nu\, *} ]
&=& 
 4 g^2_{R_a} \left\{ -g^{\mu\nu} (k_1 \cdot k_2) + 
k_1^\mu k_2^\nu + k_2^\mu k_1^\nu 
+ i \epsilon^{\mu\nu\rho\sigma} k_{1 \,\rho} k_{2 \, \sigma}
\right\} 
\nonumber \\ 
&+& 
 4 g^2_{L_a} \left\{ -g^{\mu\nu} (k_1 \cdot k_2) + 
k_1^\mu k_2^\nu + k_2^\mu k_1^\nu 
- i \epsilon^{\mu\nu\rho\sigma}  k_{1 \,\rho} k_{2 \, \sigma}
\right\} 
\nonumber \\ 
 &+&  4 ( g_{R_a} g_{L_a} + g_{L_a} g_{R_a} )
      \left\{  +g^{\mu\nu} (m_1 m_2)  \right\}
                \label{eq:omgTr}
\end{eqnarray}
where $\{ g_{R_a}, g_{L_a} \}$ are the couplings of the $a$-type  parton to the
boson,  and the on-mass-shell delta function is given by \Eq{hqkMsShll}.  

This expression for $\omega^\mu_\nu$ can be used in two ways: (i) it can be
substituted into the general factorization theorem formula, \Eq{FactThm:Wmunu},
and then contracted with $L^\nu_\mu$ to yield leading order cross-sections
directly, \cf \Eq{dsigLW}; or (ii) it can be used to calculate the helicity
structure functions through \Eq{PtnStrFun-Hel} and \Eq{FactThm-Hel} before
substituting into the general cross-section formula \Eq{dSigma-Hel}. We shall do
both, and demonstrate the consistency of the two approaches.  Although at 
leading order, these two methods are comparable in the ease of use, the
helicity approach provides a  more efficient way of calculating  higher
orders.  It also provides additional insight on the stucture of the physical
amplitudes, as we will discuss.

We begin with the helicity approach  using 
\begin{eqnarray}
\omega_{\lambda} &=& 
{1 \over 4 \pi} \ 2 \pi \delta_+(k_2^2-m_2^2) \ 
\overline{\sum_{\sigma_1 , \sigma_2}} \ 
{J^*}^{\sigma_1 \sigma_2}_{\lambda}\ 
{J  }_{\sigma_1 \sigma_2}^{\lambda}
\nonumber \\ [5pt] 
&=&
\epsilon_{\la}^{\mu \, *}  (k,q) \ 
\omega_{\mu\nu} (k,q)  \ 
\epsilon_{\la}^{\nu}  (k,q)   
\qquad {\rm no \ sum \ on } \ \lambda 
\end{eqnarray}
and \Eqs{loOmg}{omgTr} above for $\omega_{\mu\nu} (k,q)$, the helicity structure
functions at the parton level can be evaluated.  
We obtain\rlap,\footnote{
 Note that we have used $\Omega^{RL}=\Omega^{LR}$ to simplify \Eq{omgHel}, and 
$\omega$ is symmetric under $\Omega^{RL} \rightarrow \Omega^{LR}$. }
\begin{eqnarray}
\omega_{\lambda}  &=&
{ \delta\left(  {\xi \over \chi } - 1 \right) } \ 
(  g^2_{R_a}       \ {\Omega}_{\lambda}^{RR} + 
 2 g_{R_a} g_{L_a} \ {\Omega}_{\lambda}^{RL} + 
   g^2_{L_a}       \ {\Omega}_{\lambda}^{LL}    )
\nonumber \\          \label{eq:omgHel}
\end{eqnarray}
where the superscripts $(R, L)$ refer to right-handed and left-handed chiral
couplings at the hadron vertices,
and the ${\Omega}$'s are given in \Tbl{loOmg}.

\begin{table}[tbh]
\bigskip
\[
\begin{array}{||rcl||c|c|c||}\hline\hline
\ttstrut
&{\Omega}_{\lambda}^{\chi\, \chi'}&  & 
\chi\chi' = RR & \chi\chi' = RL= LR & \chi\chi' = LL  
\\ \hline
\tstrut
& & & g^2_{R_a}   &  2 g_{R_a} g_{L_a} &  g^2_{L_a}  
\\  \hline   \hline  
\ttstrut
\lambda &=& + & 
{Q^2+m_1^2+m_2^2+\Delta \over \Delta} & 
{-2m_1 m_2 \over \Delta}   &
{Q^2+m_1^2+m_2^2-\Delta \over \Delta}  
\\ \hline
\ttstrut
\lambda &=& 0 & 
{ (m_1^2+m_2^2) + (m_1^2-m_2^2)^2/Q^2 \over \Delta}  &
{+2m_1 m_2 \over \Delta}   &
{ (m_1^2+m_2^2) + (m_1^2-m_2^2)^2/Q^2 \over \Delta}
\\ \hline
\ttstrut
\lambda &=& - & 
{Q^2+m_1^2+m_2^2-\Delta \over \Delta}  &
{-2m_1 m_2 \over \Delta}   &
{Q^2+m_1^2+m_2^2+\Delta \over \Delta}
\\ \hline \hline
\end{array} 
\]
  \caption{ 
The helicity amplitudes for the leading-order process 
$\ell_1  + k_1(m_1) \rightarrow \ell_2  + k_2(m_2) $,
with $\Delta = \Delta[-Q^2,m_1^2,m_2^2]$. 
\label{tab:loOmg}  
}
\end{table}

The partonic helicity structure functions $\{ {\omega}_\la \}$ exhibit many
physically interesting features which are obscured in the conventional Dirac
trace method. 
For example, there are obvious symmetries under $g_{R_a}\leftrightarrow
g_{L_a}$ when the vector boson helicity is flipped. Additionally, there is
a clear  order of magnitude separation of the amplitudes when $m_{1,2}^2/Q^2$
become small (high energy limit)---all the longitudinal structure functions, as
well as the mixed chirality ones, become of ${\cal O}(m_{1,2}^2/Q^2)$.

Because of the direct relationship between the hadronic helicity structure
functions \{$F_\la$\} to the partonic helicity structure
functions\{$\omega_\la$\}, the \{$F_\la$\} functions are essentially given by the
expressions above multiplied by the relevant parton distribution functions
evaluated at $\xi = \chi$ (due to the delta function in \Eq{omgHel}).
Substituting these expressions in the general formula for $L\cdot W$,
\Eq{WdotL-H}, we obtain:
\begin{eqnarray}
L \cdot W  = q(\xi)  \   \otimes \ {8\over n_\ell}  \                 
\Bigg\{ & g_{R\, \ell}^2 & 
\left[
 \omega_+ \left( {1+\cosh\psi \over 2} \right)^2 
+\omega_0 \left( {-\sinh\psi \over \sqrt{2}} \right)^2 
+\omega_- \left( {1-\cosh\psi \over 2} \right)^2 
\right] 
\nonumber                                                              
\\+ & g_{L\, \ell}^2 &
\left[
 \omega_+ \left( {1-\cosh\psi \over 2} \right)^2 
+\omega_0 \left( {+\sinh\psi \over \sqrt{2}} \right)^2 
+\omega_- \left( {1+\cosh\psi \over 2} \right)^2 
\right] 
\Bigg\}
\nonumber \\
\label{eq:wdotL-H}
\end{eqnarray}
with $\{\omega_+, \omega_0, \omega_-  \}$ given by \Eq{omgHel}. 
 The corresponding results for the anti-quark process is obtained by the substitution
$g_{R_a} \leftrightarrow g_{L_a}$.

Alternately, we can compute this in the tensor representation by 
contracting  $\omega^\mu_\nu$ with  $L^\nu_\mu$, \Eq{apLepTen}, to obtain:
\begin{eqnarray}
{L \cdot \omega} &=& {1\over n_\ell} \ 
{2^6 \over Q^2} \ 
{\delta\left(  {\xi\over \chi} - 1 \right)
  \over   \Delta[-Q^2,m_1^2,m_2^2]  }  \  
\left\{ 
\begin{array}{rrl}
  &(g^2_{R_a} g^2_{R\, \ell} + g^2_{L_a} g^2_{L\, \ell} ) 
   &(k_1\cdot \ell_1)(k_2\cdot\ell_2) \\  [5pt]
+ &(g^2_{R_a} g^2_{L\, \ell} + g^2_{L_a} g^2_{R\, \ell} ) 
   &(k_1\cdot \ell_2)(k_2\cdot\ell_1) \\ [5pt]
- &g_{R_a} g_{L_a} (g^2_{R\, \ell} +g^2_{L\, \ell} ) 
   &(m_1 m_2) (\ell_1 \cdot\ell_2) 
\end{array}
\right\}
\nonumber\\       \label{eq:LdotOmgGen}
\end{eqnarray}
Applying the convolution integral and inserting the scalar products between
lepton and quark momenta  derived in  \App{lepDotHad} into \Eq{LdotOmgGen} 
leads to:
\begin{eqnarray}
{L \cdot W} &=& {1\over n_\ell} \ 
{2^6 \over Q^2} \ 
{ q(\chi)  \over   \Delta[-Q^2,m_1^2,m_2^2]  }  \ \times 
\nonumber \\ [10pt] 
&&  
\left\{ 
\begin{array}{rrl}
  &(g^2_{R_a} g^2_{R\, \ell} + g^2_{L_a} g^2_{L\, \ell} ) 
   & 
( Q^2 \chi^2 d_- + m_1^2  \eta^2 d_+ + \chi \eta Q^2 )
( Q^2 \chi^2 d_+ + m_1^2  \eta^2 d_- )
/ (  2^2 \eta^2 \chi^2  )
\\  [5pt]
+ &(g^2_{R_a} g^2_{L\, \ell} + g^2_{L_a} g^2_{R\, \ell} ) 
   & 
( Q^2  \chi^2 d_+ + m_1^2 \eta^2 d_- - \chi \eta Q^2 )
( Q^2 \chi^2 d_- + m_1^2  \eta^2 d_+ )
/ (  2^2 \eta^2 \chi^2  )
\\ [5pt]
- &g_{R_a} g_{L_a} (g^2_{R\, \ell} +g^2_{L\, \ell} ) 
   &(m_1 m_2) \ Q^2/2  
\end{array}
\right\}
\nonumber\\       \label{eq:LdotWloGen}
\end{eqnarray}
where $d_\pm=(\cosh\psi \pm 1)/2$ are elements of the $d^1(\psi)$ matrix.
 A special case of these results -- charm production in neutrino scattering --
is discussed in \Sec{LO}.
  
Although it is far from obvious, \Eq{wdotL-H} and \Eq{LdotWloGen}
are in fact identical (as some tedious algebra will prove). The difference in
appearance is simply that the helicity approach exploits the symmetries of the
problem; hence, these symmetries are manifest in the
final representation of the cross section,  \Eq{wdotL-H}.




\section*{Acknowledgement}

The authors would like to thank 
Andrew Bazarko,  
Raymond Brock,  
John Collins, 
Sanjib Mishra, 
Michael Shaevitz,
and 
Davison Soper
for useful discussions.
This work is partially supported by 
the National Science Foundation under Grant No.PHY89-05161, 
the U.S. Department of Energy Contract No. DE-FG05-92ER-40722,
and by 
the Texas National Research Laboratory Commission. 
M.A. and F.O. also thank the Lightner-Sams Foundation for support. 
F.O. is supported in part by an SSC Fellowship.


 \newcommand{\bibtit}[1]{}  \newcommand{\bibit}[1]{\bibitem{#1}}




\end{document}